\documentclass[9pt, twocolumn]{pnas-jcm}
\templatetype{pnasresearcharticle} % Choose template 
\usepackage{times}
\usepackage[printwatermark]{xwatermark}
\usepackage{graphicx}
\usepackage{lipsum}

\usepackage{times}
\usepackage{amsmath}
\usepackage{graphicx}

\newcommand{\jcm}[1]{}
\newcommand{\dir}[1]{#1}

\newcommand{\rep}{\mathrm{rep}}
\newcommand{\figScale}{.5}
\newcommand{\figS}{.95}

\newcommand{\dem}{\mathrm{dem}}

\renewcommand{\tt}{\mathrm{tot}}
\newcommand{\dd}{\mathrm{dem}}
\newcommand{\rr}{\mathrm{rep}}

\newcommand{\R}{\mathbf{R}}

\newcommand{\pr}{\mathbf{P}}

\title{Evaluating Partisan Gerrymandering in Wisconsin}
\date{September 5, 2017}
\author[a,b]{Gregory Herschlag}
\author[a]{Robert Ravier} 
\author[a,c]{Jonathan C. Mattingly}
\affil[a]{Department of Mathematics, Duke University, Durham NC 27708}
\affil[b]{Department of Biomedical Engineering, Duke University, Durham NC 27708}
\affil[c]{Department of Statistical Science, Duke University, Durham NC 27708}

\leadauthor{Herschlag} 

\begin{abstract}
  We examine the extent of gerrymandering for the 2010 General 
  Assembly district map of Wisconsin. We find that there is
  substantial variability in the election outcome depending on what
  maps are used.  We also found robust 
  evidence that the district maps are highly gerrymandered and that 
  this gerrymandering likely altered the partisan make up of the Wisconsin General 
  Assembly in some elections.  
Compared to the distribution of possible redistricting plans for the General Assembly, 
Wisconsin's chosen plan is an outlier in that it yields results that are highly skewed 
to the Republicans when the statewide proportion of Democratic votes comprises 
more than 50-52\% of the overall vote (with the precise threshold depending on the 
election considered).  Wisconsin's plan 
acts to preserve the Republican majority by 
  providing extra Republican seats even when the Democratic vote increases 
  into the range when the balance of power would shift for the vast majority of redistricting plans.
\end{abstract}

\begin{document} 

\maketitle 

\thispagestyle{firststyle}
\ifthenelse{\boolean{shortarticle}}{\ifthenelse{\boolean{singlecolumn}}{\abscontentformatted}{\abscontent}}{}

\dropcap{W}e generate an ensemble of 19,184 redistricting plans
drawn from a distribution placed on redistricting plans of the state of
Wisconsin.  The probability distribution used is concentrated on
redistricting plans that satisfy design criteria laid out in the
Wisconsin constitution, statutes, and relevant court cases:
compactness and contiguity of districts, equal partition of votes, resistance to
splitting counties across districts, and compliance with the Voting
Rights Act (VRA). With the possible exception of satisfying the VRA, none of
these design criteria have any partisan tilt.

We explore three basic questions: the variability of elections results
across redistricting plans; the degree to which the Wisconsin Act 43
is typical or an outlier with respect to its partisan bias; and lastly, the structural source of any bias.

Our approach has a number of inherent advantages. We do not presume
any notion of proportional representation based on statewide vote
counts. By sampling, we are able to factor in the inherent
geopolitical structure of the state such as the concentration of
Democrats in urban areas or the existence of geographic elements that constrain
redistricting plans. Such features might produce basic asymmetries in the
number of representatives elected as a function of the statewide votes. We
make no symmetry assumptions and our methods naturally adapt to the
geometry of population distributions of the state.

In Section~\ref{sec:variablity}, we discuss how the
election results may vary depending on the restricting maps used. In 
Sections~\ref{sec:sitWI43}~and~\ref{sec:expos-geop-struct}, we explore the geopolitical 
structure of Wisconsin and give graphical  aids for understanding and 
detecting gerrymandering. In Section~\ref{sec:sumary-statistics}, we 
explore a number of summary statistics which quantify the understanding 
and insights developed in  Sections~~\ref{sec:sitWI43}~and~\ref{sec:expos-geop-struct}. 

We find that the Wisconsin redistricting plan is highly gerrymandered
and less representative 
than at least 99\% of all plans in our ensemble and shows more
Republican bias than  over 99\% of the plans.
The
gerrymandering results are stable over a number of different sets of
votes and years. These
results are summarized in Tables~\ref{tab:index} and~\ref{tab:sumstat}
in Section~\ref{sec:sumary-statistics}.
 These results further suggest that the election outcomes produced by the Wisconsin
 maps systematically become less
representative of our ensemble as the overall percentage of Republican
votes decreases to 50\% and below.  This is further supported by
graphical analysis in
Figures~\ref{fig:allRealData}--\ref{fig:boxPlots} in
Section~\ref{sec:expos-geop-struct}.  The Act 43 map acts as a kind of
firewall, keeping a Republican Assembly majority in place even as
voter preferences becomes increasingly Democratic. In
Figure~\ref{fig:histevenShift}, we show that there is a fundamental
asymmetry in the geopolitical structure of Wisconsin as most of the
redistricting maps in our ensemble require less than 50\% of the votes
to be Republican to make equal the chance of either party being in the
legislative majority. Nonetheless, the Act 43 map is again a clear
outlier in favor of the Republicans as it  requires a much lower
percentage of Republican votes to produce an equal chance of having
a legislative majority.

In Section~\ref{sec:generatingEnsemble}, we discribe how the ensemble
is generated by Markov Chain Monte Carlo. In
Section~\ref{sec:robustness-results}, we give evidence that our
results are robust and that the algorithm is sufficiently
converged. In particular we show that the our all of our results
remain unchanged when a
larger ensemble of 84,500 redistricting plans is used.
 In Sections~\ref{sec:interp} and~\ref{sec:adjust}, we make some
 technical comments about data curation.

This report continues  our work started in
\cite{MattinglyVaughn2014,beyondGerry,2017arXiv170403360B}. It is
related to other works on sampling and computation in the
redistricting context
\cite{thoreson1967computers, gearhart1969legislative,Wu15,Chen15, Liu16,fifield2015,Chikina_Frieze_Pegden_2017,wang16:_three_tests_pract_evaluat_partis_gerry}. In
particular, the recent papers \cite{chen17,2017arXiv170809852C} apply sampling to the Wisconsin
redistricting setting that we consider here.

\section{The Inherent Variability of Election Results}
\label{sec:variablity}
For each redistricting plan in the ensemble, the outcome of the election is
computed using votes from either the Wisconsin General Assembly elections
from 2012 (denoted WSA12), from 2014 (denoted WSA16) and from 2016
(denoted WSA16). In all cases, the actual votes were used at the ward
level. However, the existence of unopposed races necessitated
interpolating the data using votes from other elections in a number of
wards: 27\% in WSA12, 46\% in WSA14, and 49\% in WSA16. The details of
this interpolation are given in Section~\ref{sec:interp}, but the vote
counts are
based on actual Wisconsin election data in the years given.
%

%Hist_VRA_WSA12H_nolab

\begin{figure}[h]
  \centering
\includegraphics[width=\figScale\linewidth]{\dir{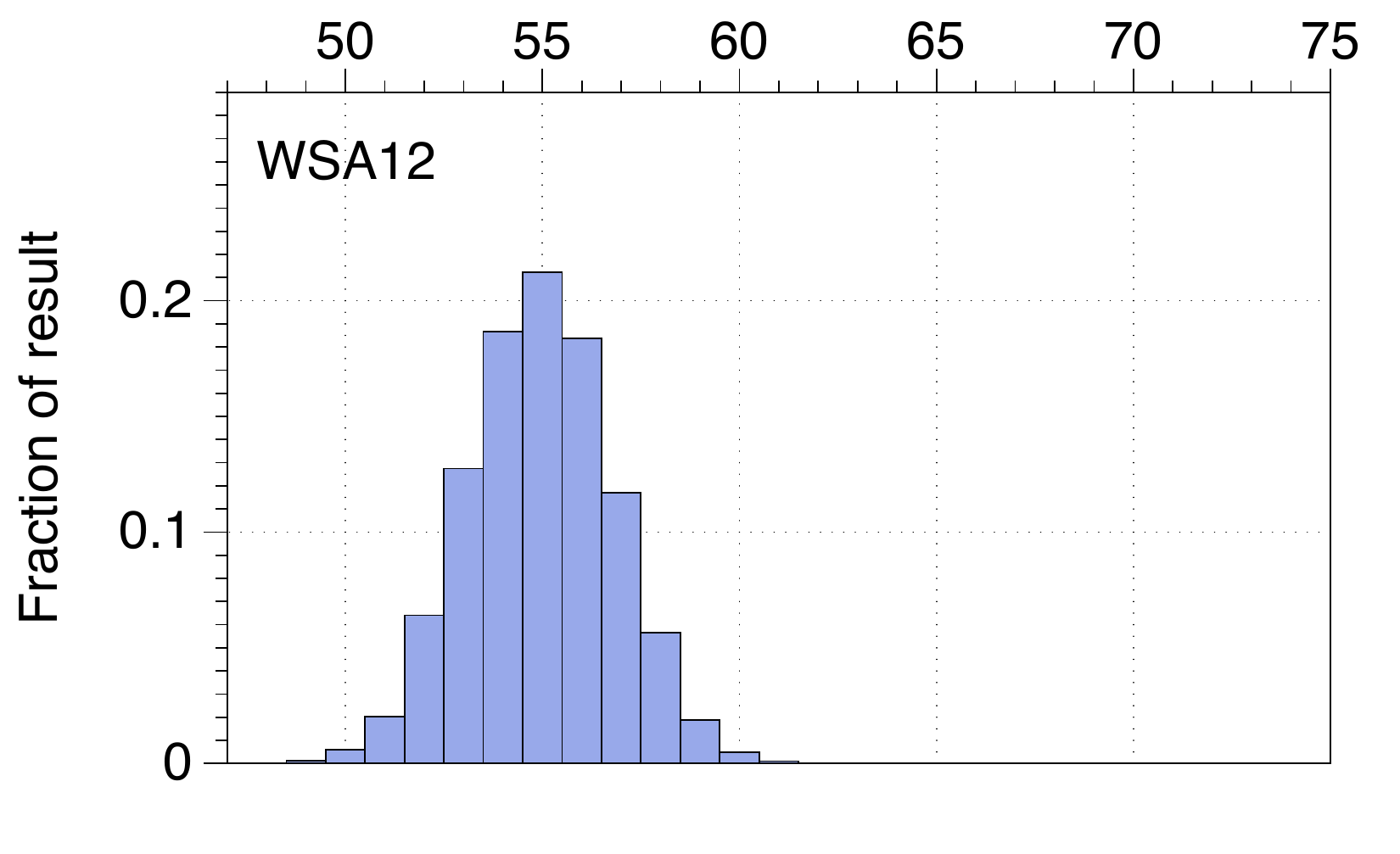}}
\includegraphics[width=\figScale\linewidth]{\dir{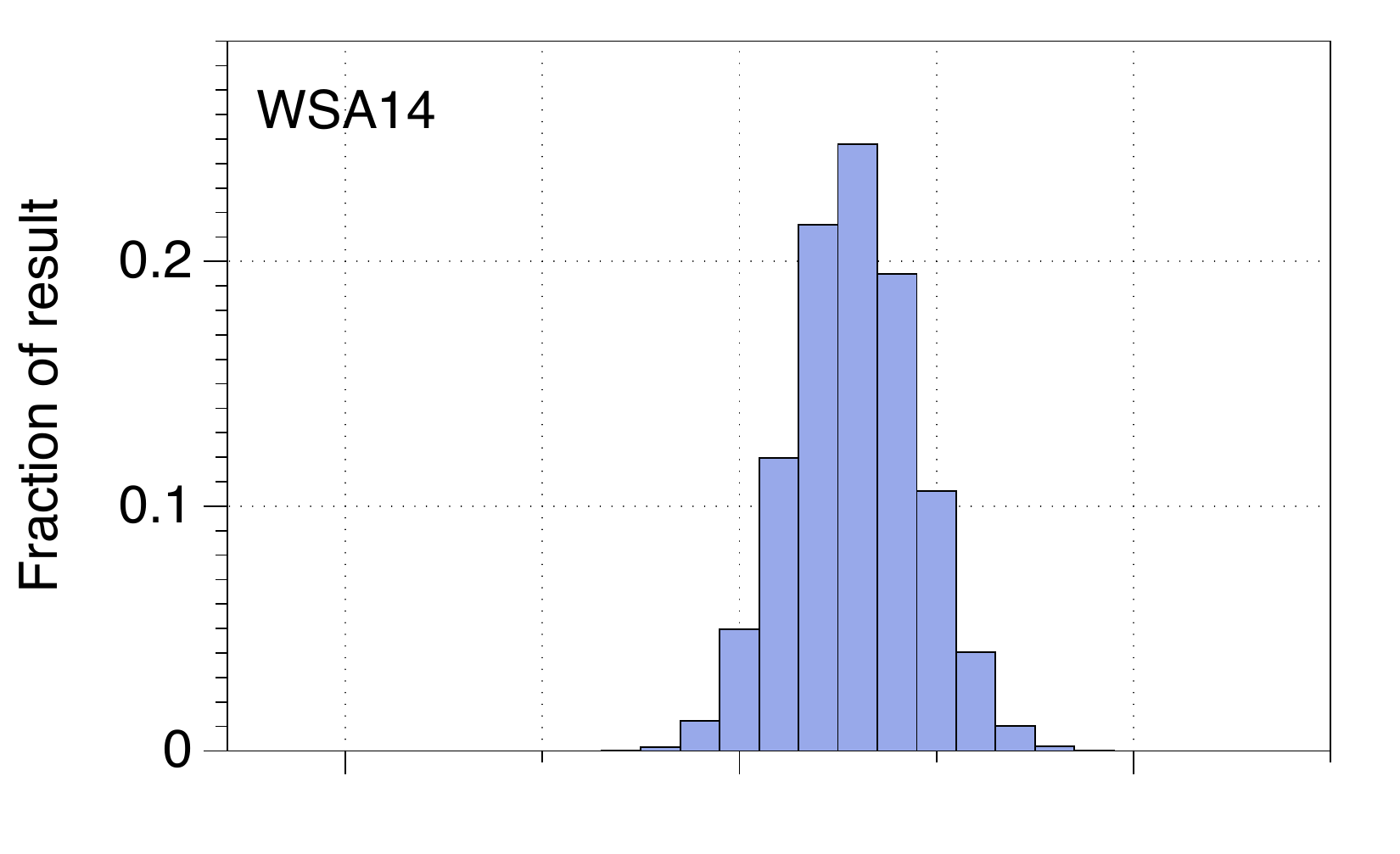}}
\includegraphics[width=\figScale\linewidth]{\dir{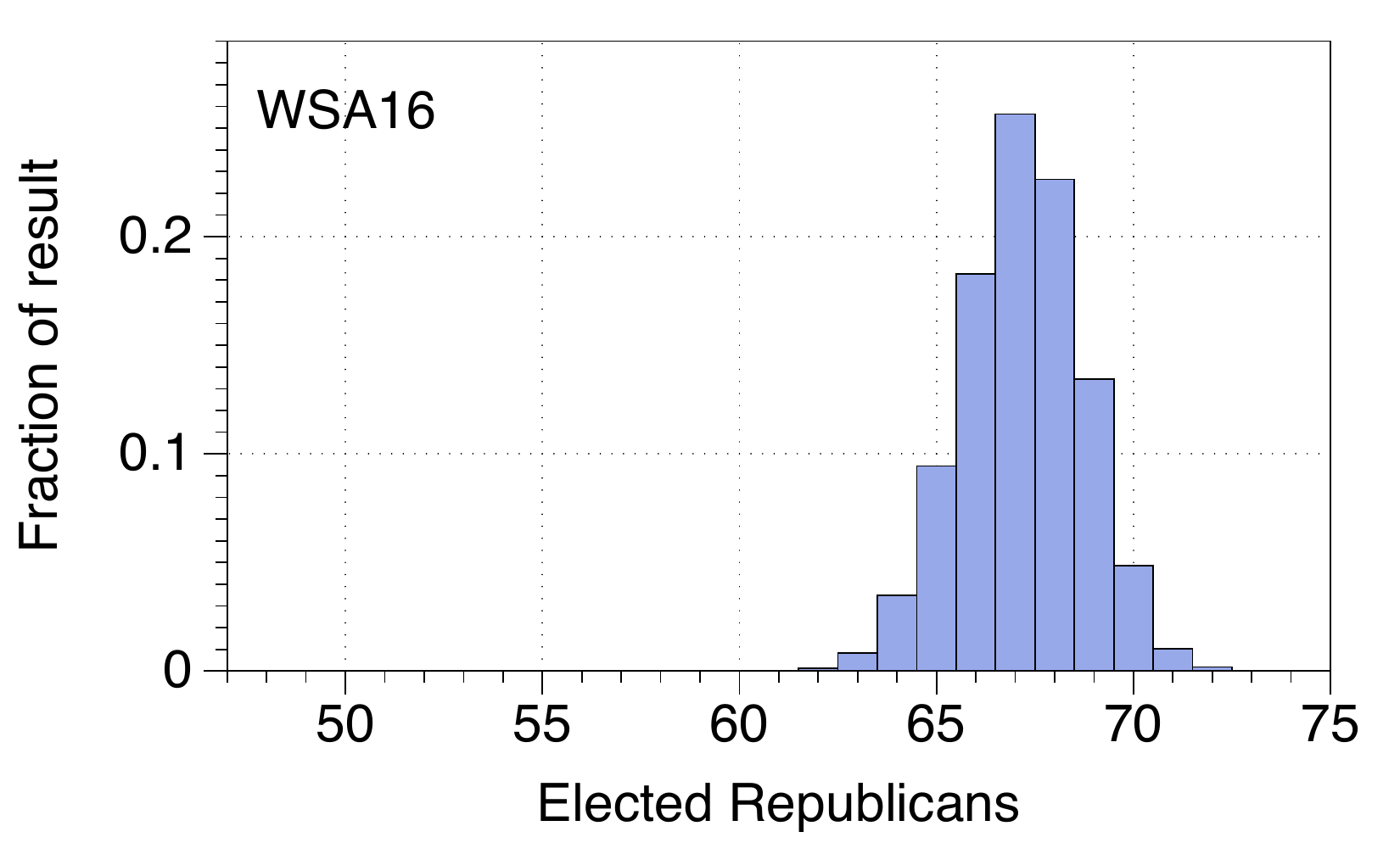}}
\caption{Distribution of election outcomes in the ensemble of 19,184 redistricting plans, 
  interpolated for the WSA12, 
  WSA14, and WSA16 election data.  With a fixed number and location of votes, the outcome of the election varies based on the choice of redistricting plan chosen. }
\label{fig:hist}  
\end{figure}

Figure~\ref{fig:hist} shows the frequency of different election
outcomes in our ensemble using the votes from WSA12, WSA14, and WSA16.
Across the redistricting plans for the 99-seat Wisconsin General
Assembly, the expected number of seats won by Republicans was
typically concentrated within a range of 3-5 seats. However, a small
proportion of redistricting plans are outliers, which extend the range
to as much as 10 seats. This wide range of possible outcomes shows
that the state's choice of redistricting plan can have an effect on
the same order as the typical changes in the popular vote across
elections (e.g. a swing of 60 to 64 elected Republicans from 2012 to
2016 in the Wisconsin General Assembly).  The fact that the different
redistricting plans in the ensemble give such different results speaks
to the need for a concept of acceptable redistricting, lest the
state's redistricting exercise become as, or more important than, the
democratic expression of voters.

While the precise definition of a \textit{typical result} may be
debatable, it is clear that some extreme ranges clearly represent
anomalous behavior: the results  should be labeled as
outliers. The view that some points would clearly be labeled as
outliers is the starting point for our analysis.

\section{Situating the Wisconsin  Act 43 Redistricting in the Ensemble}
 \label{sec:sitWI43}
\begin{figure}[h]
  \centering
\includegraphics[width=\figScale\linewidth]{\dir{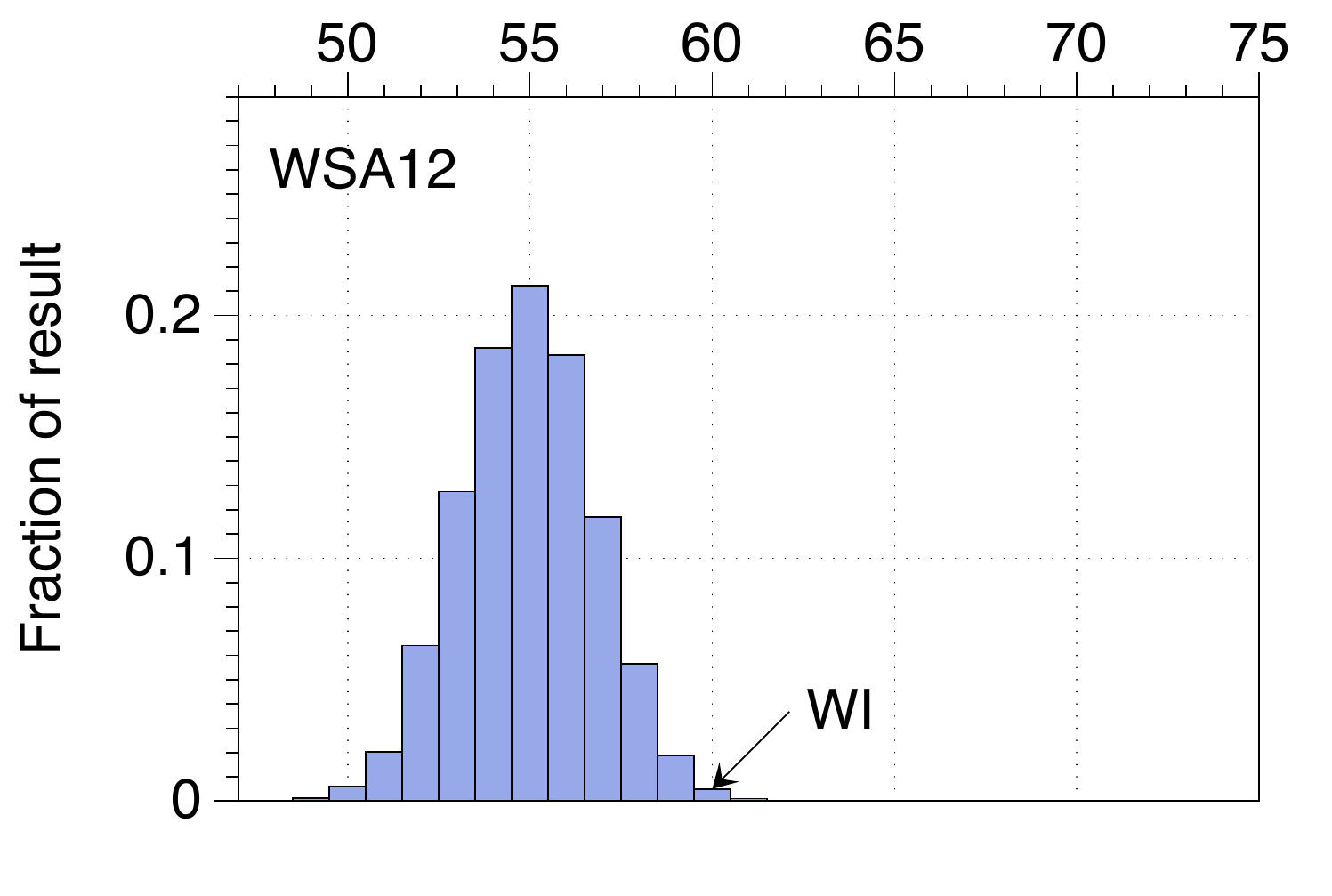}}
\includegraphics[width=\figScale\linewidth]{\dir{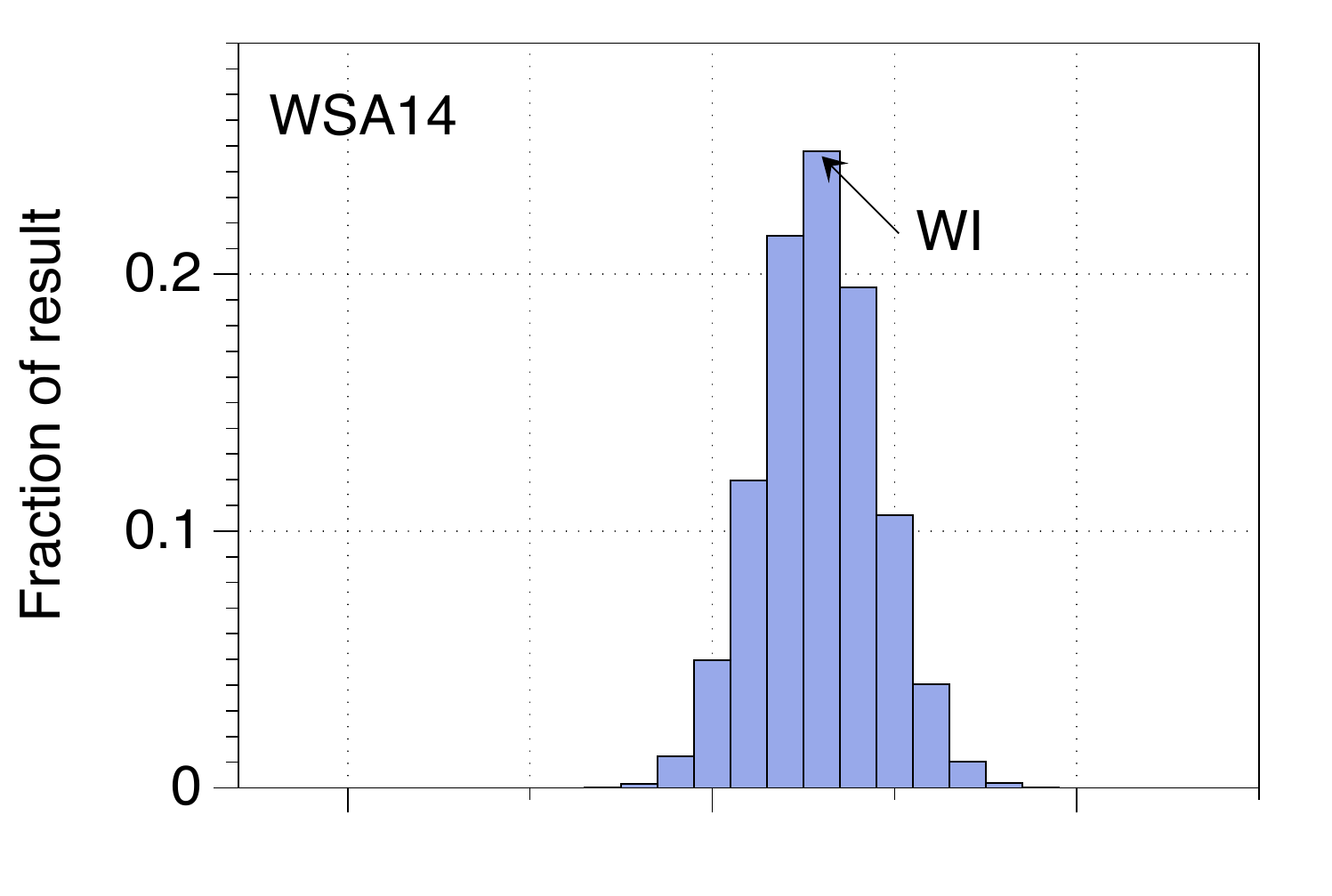}}
\includegraphics[width=\figScale\linewidth]{\dir{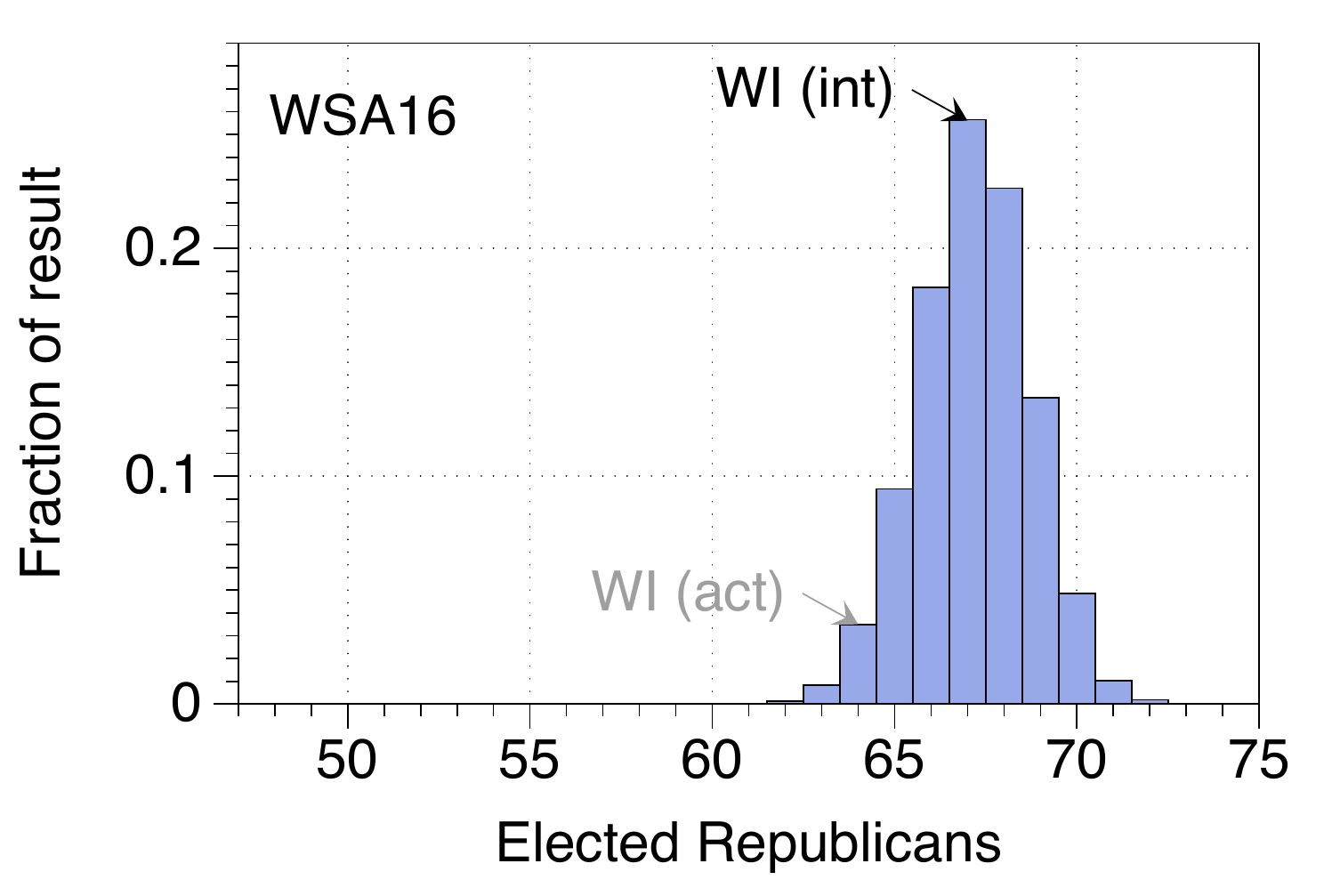}}
\caption{Distribution of election outcomes in the ensemble of 19,184 redistricting plans, 
  interpolated for the WSA12, 
  WSA14, and WSA16 election data. The outcome using the Wisconsin Act 43 redistricting
  is marked with ``WI''. For the WSA16 outcome, there were three unopposed Democrats that ran in districts that voted more Republican across the interpolated data; thus we have marked the actual result (64), along with the interpolated result (67).  }
\label{fig:histlab}  
\end{figure}

We now turn to situating the actual redistricting plan established by Act
43 of the 2011 Wisconsin General Assembly within our ensemble of $19,184$
redistricting plans. This was the redistricting plan actually used in the WSA12,
WSA14, and WSA16 elections.  The annotation
``WI'' on each plot in Figure~\ref{fig:histlab} indicates the number of
seats produced by this redistricting. We note that the use of
our modified election data in 2012 and 2014, which interpolates the missing data caused
by unopposed races, does not change the balance of power. However, in 2016 the results of the
actual election differed from those our interpolated vote
data produces. The actual results had three fewer Republican seats than
the interpolated results would have had, due to unopposed races in which Democrats 
ran unopposed in districts that tended to vote Republican. The number of
unopposed races was least  in 2012 with 27\%, growing to 46\%
in 2014, and then to 49\%  in 2016.

Any reasonable sense of outlier would label the Wisconsin
result in 2012 as anomalous. Yet, the actual result produced by the same map 
is well within the distribution for 2014 and 2016, in which the Republican share of the vote 
was considerably higher. As we see below, this behavior turns out to reflects an unusual 
property of Wisconsin's redistricting plan: it gives an anomalously high number of seats to 
Republicans in elections in which Democrats perform well, but a typical number of seats in 
elections in which Republicans perform well. 

To better understand this situation, we consider
the outcome that would occur if votes from a number of other elections were used as
if they had been cast for the Wisconsin General Assembly. Specifically we compare the effect of using
results from U.S. House, U.S. Senate, and
Presidential elections from 2012, 2014, and 2016 in addition to our
interpolated results for WSA12, WSA14, and WSA16.
\begin{figure}[h]
  \centering
  \includegraphics[width=.75\linewidth]{\dir{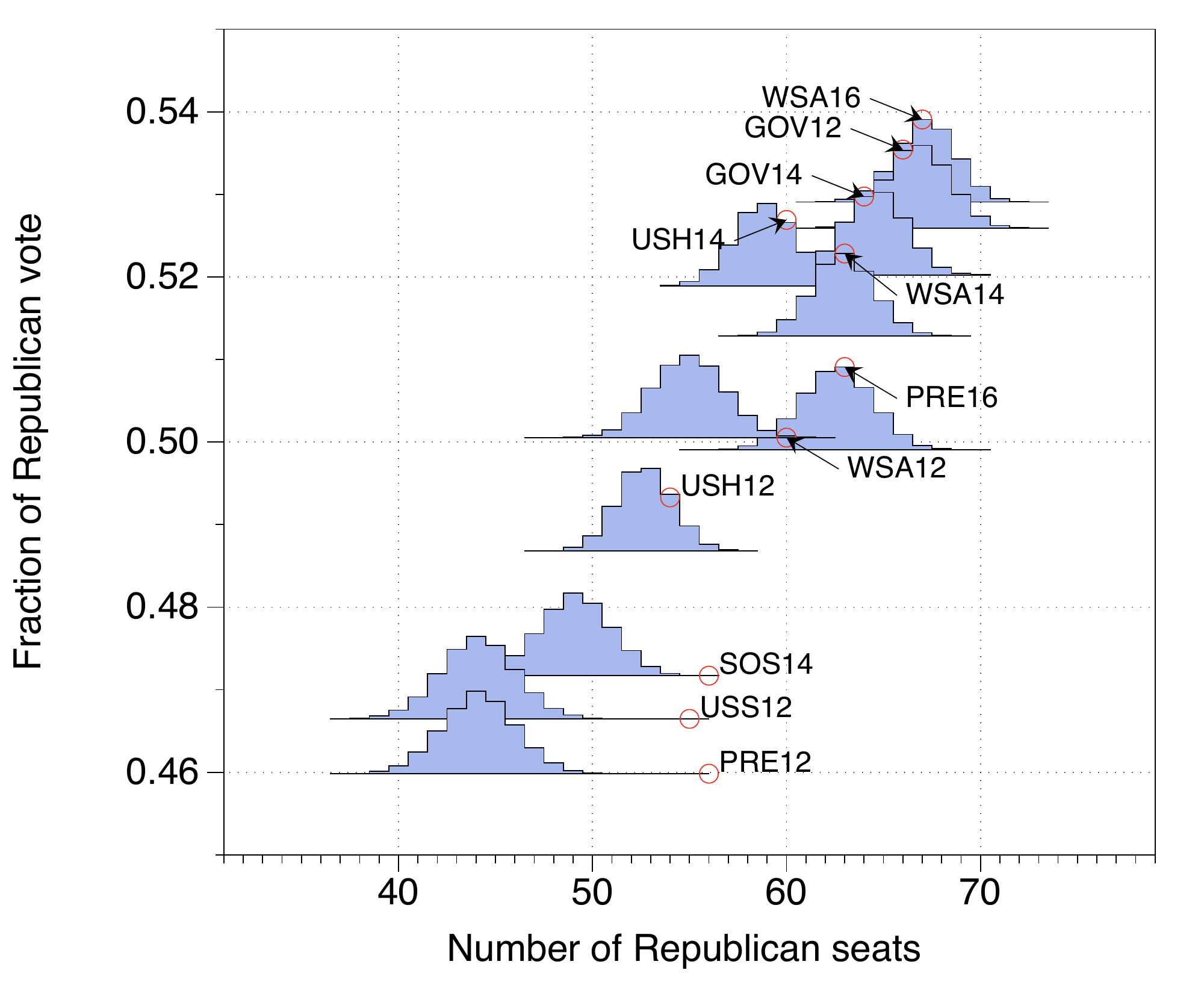}}
  \caption{A number of election seat result histograms situated on a
    larger plot by the number  Republican seats and the overall
    fraction of Republican vote.  The circles mark the outcome using
    the Wisconsin Act 43 redistricting. Election plotted: Wisconsin State
    Assembly 2012 (WSA12), Wisconsin State Assembly 2016 (WSA16),
Presidential 2012 (PRE12),
Presidential 2016 (PRE16), US House 2012 (USH12),
US House 2014 (USH14), US Senate 2012 (USS12),
US Senate 2016 (USS16), Wisconsin Secretary of State 2014 (SOS14)}
  \label{fig:allRealData}
\end{figure}
Figure~\ref{fig:allRealData} presents an interesting trend: whenever
the election would have typically produced around 55 or fewer
Republican seats, the Wisconsin plan behaves very anomalously in
the sense that it is far to the right in the histogram. In fact,
even though the expected number of Republican seats falls below 50 in one
election and the statewide percentage of Republican votes falls well below
50\% in three elections, the number of seats elected stays pinned
in the high 50s; it is almost constant despite the fact that the Republican vote
continues to fall as one moves down the plot. 

The plot shows that to determine whether the outcome of a given map
will be anomalous for a particular election, it is not enough to consider only the total vote count or expected number of seats.
The USH12 and WSA12 are similar by those two
metrics, yet the outcome in the first is typical but the second is
anomalous. This detail shows the importance of the geopolitical
structure of the votes in determining the outcome and the pitfalls of
coarse, global  measures.

\begin{figure}[h]
  \centering
 \includegraphics[width=1\linewidth]{\dir{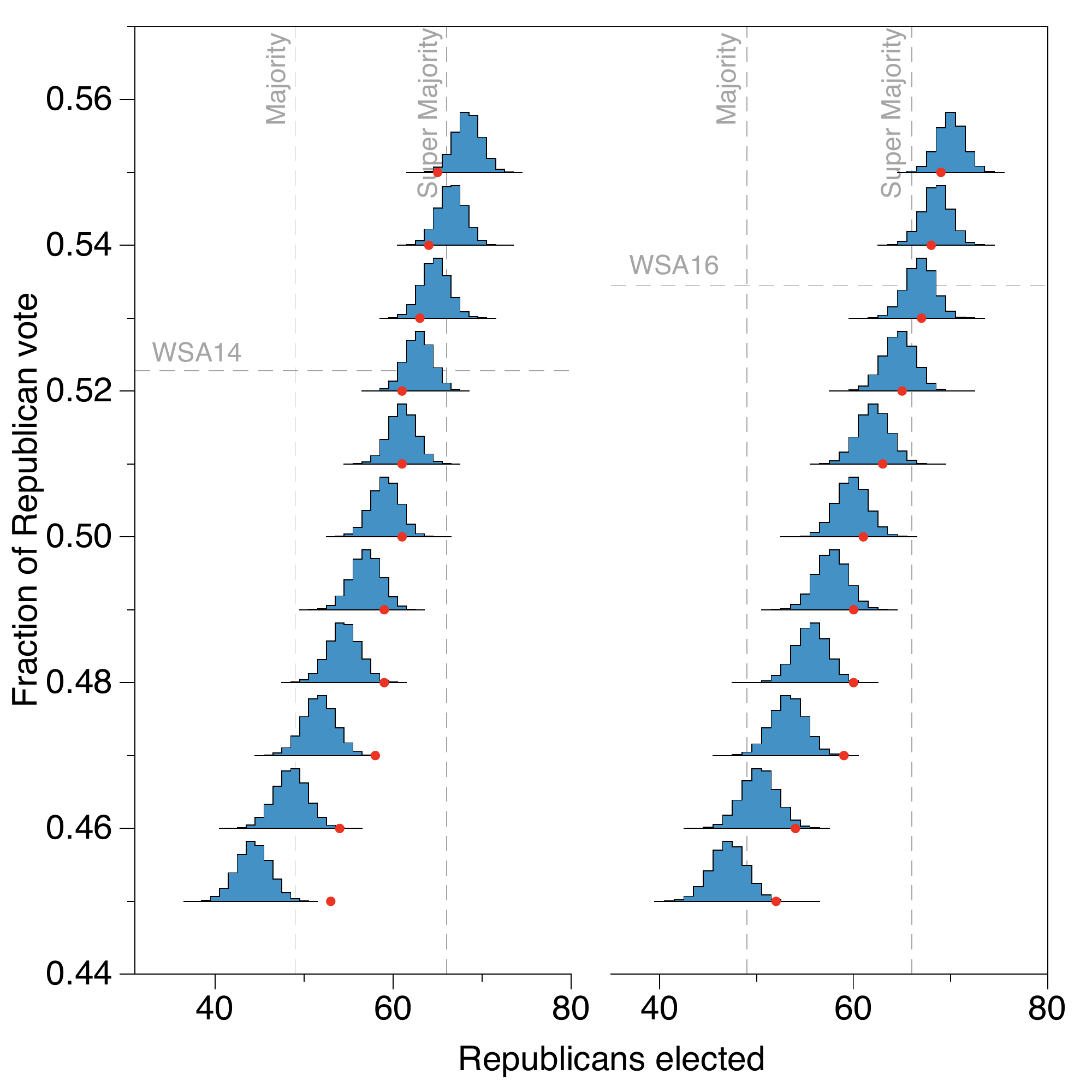}}
  \caption{The number of seats elected when the percentage in each
    district is shifted so that the global fraction of the vote for the
    Republicans ranges
    between 0.45 and 0.55. Results of WSA14(left) and WSA16(right) are
    shown. Horizontal lines mark the level of the
    original vote. Vertical lines mark the number of seats require for
  a majority and a super-majority.}
  \label{fig:shifHist}
\end{figure}

Based on these insights, we propose a method to evaluate the extent to which a state redistricting plan is an outlier with respect to its ability to protect a party from losing seats: 
 we examine the impact of shifting the proportion of votes up or down within each election examined. We shift the proportions in WSA14 and WSA16 uniformly up
and down in all districts so that the statewide Republican vote fraction varies from 45\% to 55\%.
We then plot the histograms and the election
results for each shifted vote count in Figure \ref{fig:shifHist}. 

Unlike the plots in 
Figure~\ref{fig:allRealData}, the geopolitical structure of all of
these shifted votes is identical. 
Both plots in Figure~\ref{fig:shifHist} exhibit the trend we already observed in 
Figure~\ref{fig:allRealData}. As the percentage of Republican votes 
decreases, the election results for both WSA14 and WSA16 (shown with 
red dots in Figure~\ref{fig:shifHist}) move from being representative 
(located in the center of the histograms) to being outliers (located 
in the extrema of the histograms). 

The Wisconsin
redistricting seems to create a firewall which resists Republicans 
falling below 50 seats. The effect is striking around the
mark of 60 seats where the number of Republican seats
remains constant, despite the fraction of votes   dropping from 
51\% to  48\%. 

\begin{figure}[h]
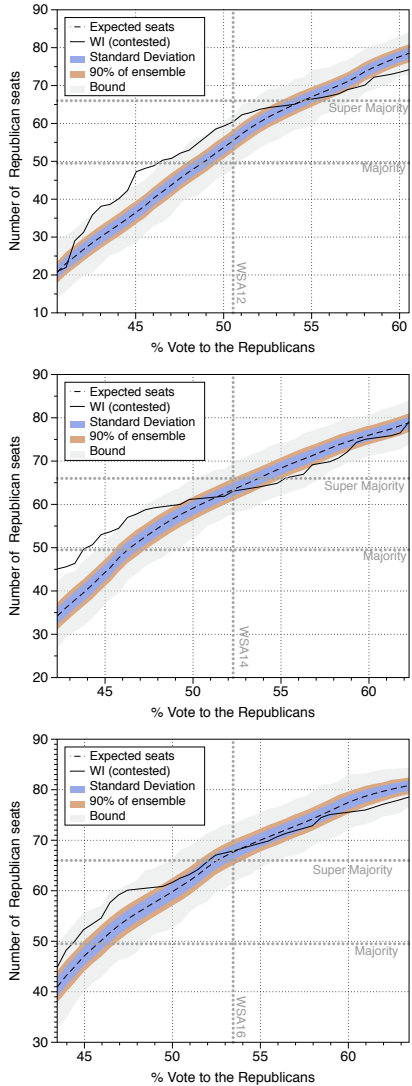

  \centering
  \includegraphics[width=.65\linewidth]{\dir{shiftEnvelopepm10WSA12}}
\includegraphics[width=.65\linewidth]{\dir{shiftEnvelopepm10WSA14}}
\includegraphics[width=.65\linewidth]{\dir{shiftEnvelopepm10WSA16}}
  \caption{Partisan composition of Wisconsin General Assemble as a 
    function of global Republican vote using
     shifted WSA12(top), WSA14(middle), WSA16(bottom) votes. 
    Vertical line indicates the actual votes in  unshifted data. Horizontal lines mark seats
    needed for majority and super-majority. Thin line shows seats in Wisconsin redistricting.}
\label{fig:contShiftPlot}
\end{figure}
Figure~\ref{fig:contShiftPlot} gives a more stylized version of
Figure~\ref{fig:shifHist}. Rather than the entire histogram, we plot the mean,
variance, a region containing 90\% of all sampled redistricting plans, and the extrema for a larger
number of finer shifts that swing the election 10 percentage points in both directions of the observed result. The fine black line gives the number of seats produced by the Wisconsin
Act 43 redistricting. Though the local geography of the votes in the
three elections is
different, each produces a clear deviation from the typical results
starting around 50\%. This deviation continues as the fraction of Republican
votes decreases. All three elections (especially WSA14 and WSA16)
show a
significant  range of Republican votes where the partisan outcome of
the election (expressed in the number of Republican seats) does not change even though the percentage of
the Republican votes  decreases substantially.

\begin{figure}[h]
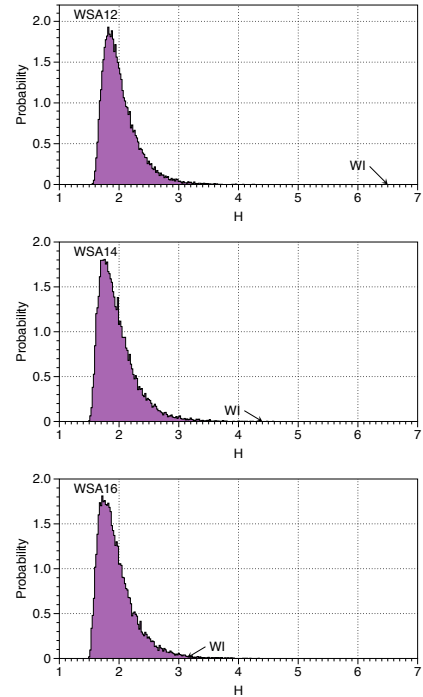

  \centering 
  \includegraphics[width=.65\linewidth]{\dir{WSA12H}}
\includegraphics[width=.65\linewidth]{\dir{WSA14H}}
\includegraphics[width=.65\linewidth]{\dir{WSA16H}}
  \caption{We plot the distribution of $H$ indices, defined in
    Section~\ref{sec:sumary-statistics}, for each set of voting data.  In all cases, we find that the Wisconsin Act 43 redistricting plan is an extreme outlier when compared with the ensemble.}
\label{fig:HindexDist}
\end{figure}

Finally, we seek to summarize the extent to which Wisconsin's redistricting plan is an outlier (compared to the ensemble of redistricting plans). 
Toward this end, we defined a statistic as follows: for each of
various shifts around an equally split election, 
we (i) calculate the extent to which a state's redistricting plan
produces results different from the results produced by the distribution of possible redistricting plans,
and (ii) take the average across the shifts. For any election, we then 
measure the extent to which a state's plan is outlier with respect to 
it statistic. (See Section~\ref{sec:sumary-statistics} for more details.)   As show in Figure~\ref{fig:HindexDist}, we find that the
Wisconsin plan is an extreme outlier.  In each of the three elections
(2012, 2014, and 2016), it is more extreme than 99\% of all possible
redistricting plans in our ensemble (See the values of $H$ in
Table~\ref{tab:sumstat} in Section~\ref{sec:sumary-statistics}). This statistic is essentially the 
log-likelihood of seeing the election outcome produced in the Wisconsin plan
averaged across the shifted elections. 

The above statistic  is symmetric in that it measures if anomalous
results favors one political party over the other, not which party. Using a second set
of statistics, we measure  if one party is
favored over the other  by the  Wisconsin plan. For each party, we measure (i) what
fraction of redistricting plans from our ensemble produce fewer
legislative seats than the state's plan, and
(ii) take the minimum of these fractions across all the shifts considered above. As
before,  for any election, we then 
measure the extent to which a state's plan is an outlier with respect to 
these statistics measuring party bias. In each of the three elections
(2012, 2014, and 2016),  the Wisconsin plan is more favorable to the
Republicans with respect to this statistic than 99\% of all possible
redistricting plans in our ensemble. In contrast,
the percentage of plans that favor the Democrats less at some shift 
 is much
smaller (only 3\%,
73\%, and 8\% in the 2012, 2014,
and 2016 elections respectively). See the values of $L_{\rep}$ and
$L_\dem$ in Table~\ref{tab:sumstat} in
Section~\ref{sec:sumary-statistics}). 
We will further see that, in the contexts in which the Act 43 map aids
Democrats, its effect is to make a large GOP majority somewhat smaller.
In Section~\ref{sec:robustness-results}, we also consider a
  complementary approach in which we assess shifts of up to $\pm7.5\%$
  centered around the outcomes of each election. Wisconsin's plan is
  again seen to be an extreme outlier.  

In the next section, we explain \emph{why} the Wisconsin plan is such an outlier by 
exploring the structure of the vote in more detail. The graphical
understanding of the structure of the vote developed in the next
section, and Figure~\ref{fig:boxPlots} in particular, is encapsulated  
in the Gerrymandering Index defined in \cite{2017arXiv170403360B}. In
Section~\ref{sec:sumary-statistics}, we explore the Gerrymandering
Index of the Wisconsin plan over a number of historical Wisconsin elections
(WSA12, WSA14, WSA16, Governor 2012, US House 2012 and 2014, US Senate
2012 and 2016, Wisconsin Secretary of State 2014, and Presidential 2012 and 2016). Again situating the result in our
ensemble, we fine that at worse 98\% of our ensemble had a better
Gerrymandering Index. For the majority of the elections considered, none
of the redistrictings in our ensemble had a worst Gerrymandering score
(See Table~\ref{tab:index} in Section~\ref{tab:index}).

\section{Exposing the  Geopolitical Structure of Wisconsin}
\label{sec:expos-geop-struct}

To understand the structure which leads to the results of the previous
section, we repeat the marginal analysis developed in
\cite{2017arXiv170403360B,beyondGerry}. Fixing a set of votes, for
each redistricting we calculate the percentage of Republican votes
and then place this vector of 99 numbers in increasing order. To gain
insight into the distribution of this 99-dimensional vector when
varied over our ensemble, we plot a box-plot for each marginal
direction. As standard in box-plots, the box contains 50\% of the
values, the outer whiskers bracket whichever is smaller -- 1.5 times the interquartile range from each quartile or the furthest outlier -- and the
central line through the box marks the median value.

\begin{figure}[ht]
  \centering
    \includegraphics[width=.9\linewidth]{\dir{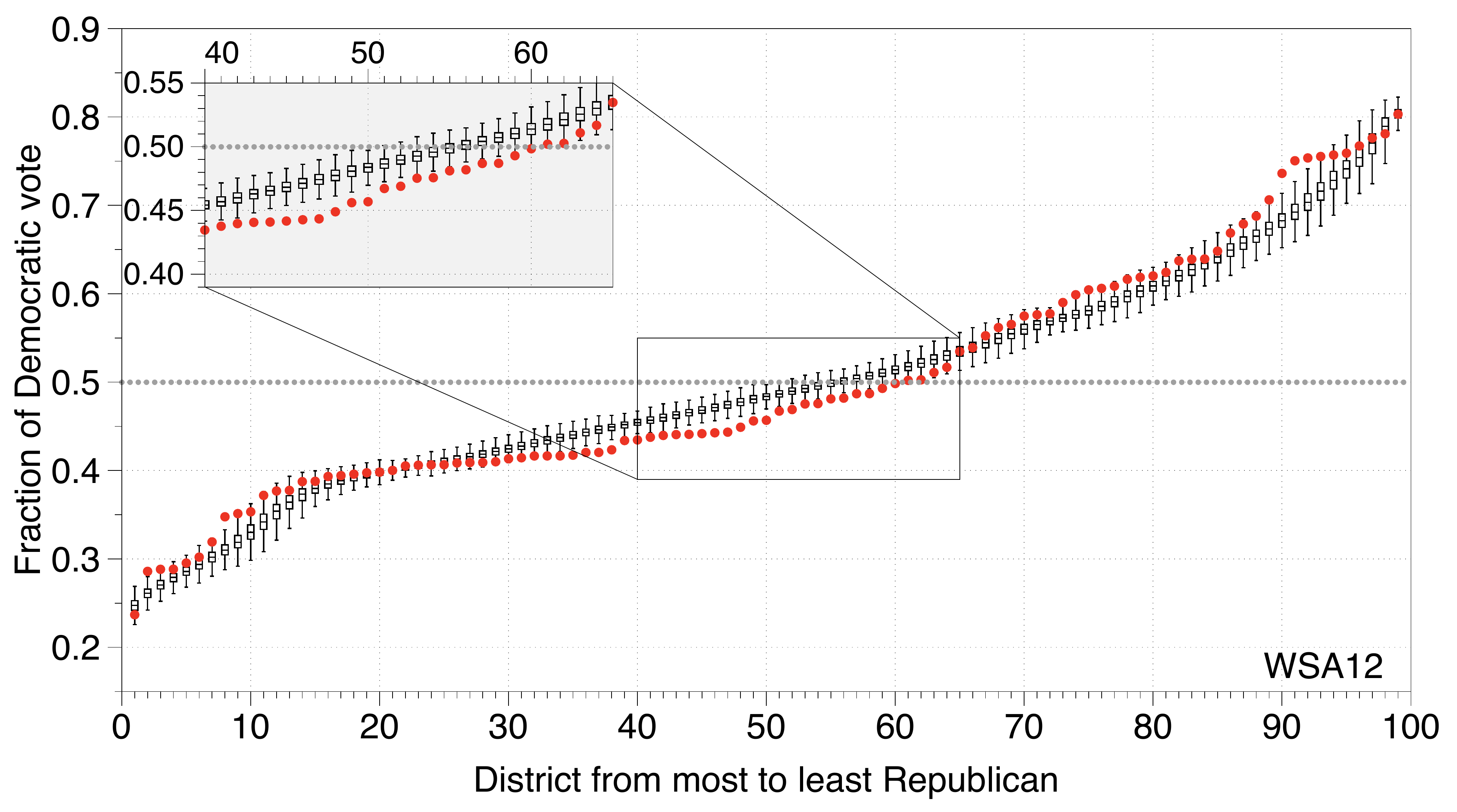}}
  \includegraphics[width=.9\linewidth]{\dir{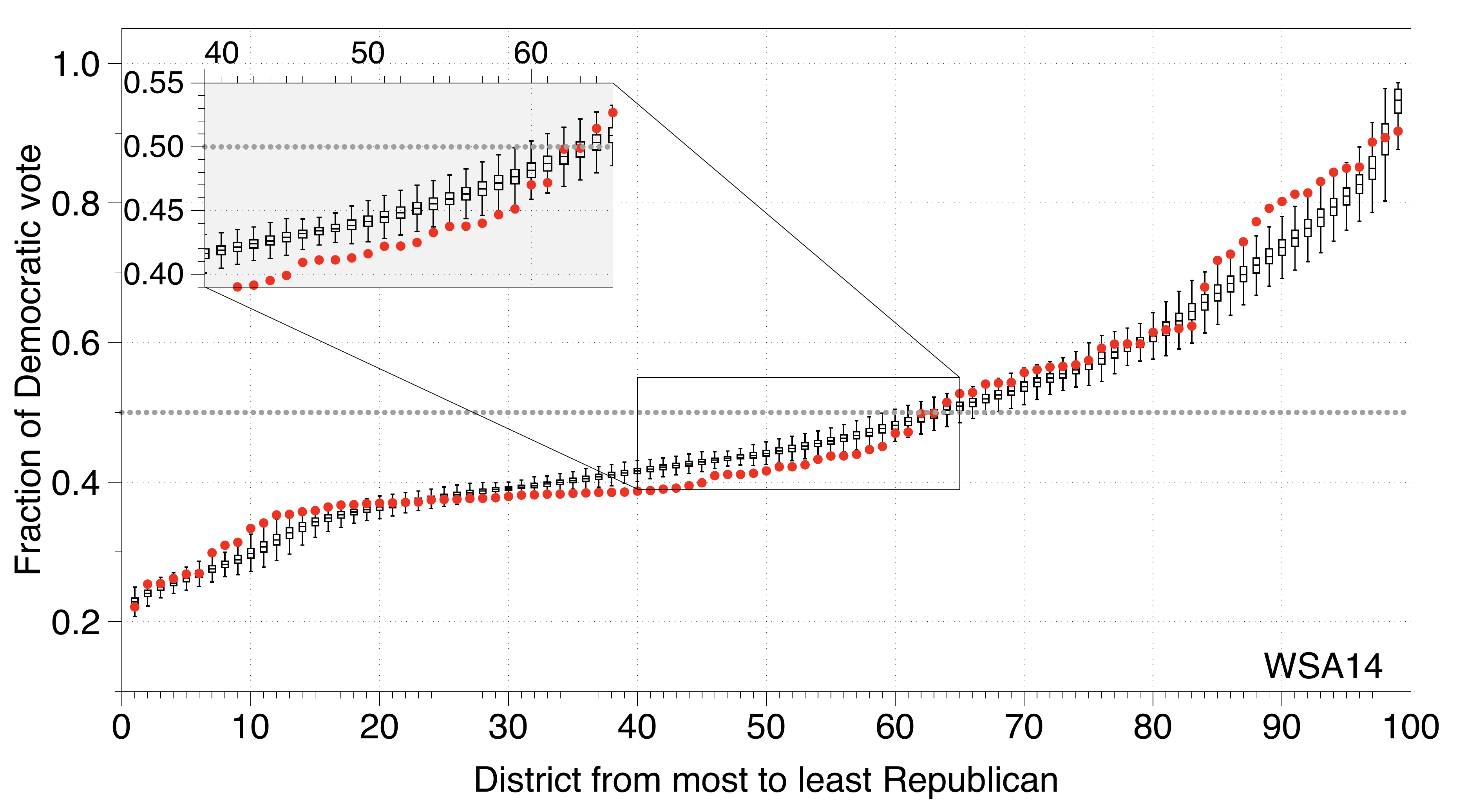}}
  \includegraphics[width=.9\linewidth]{\dir{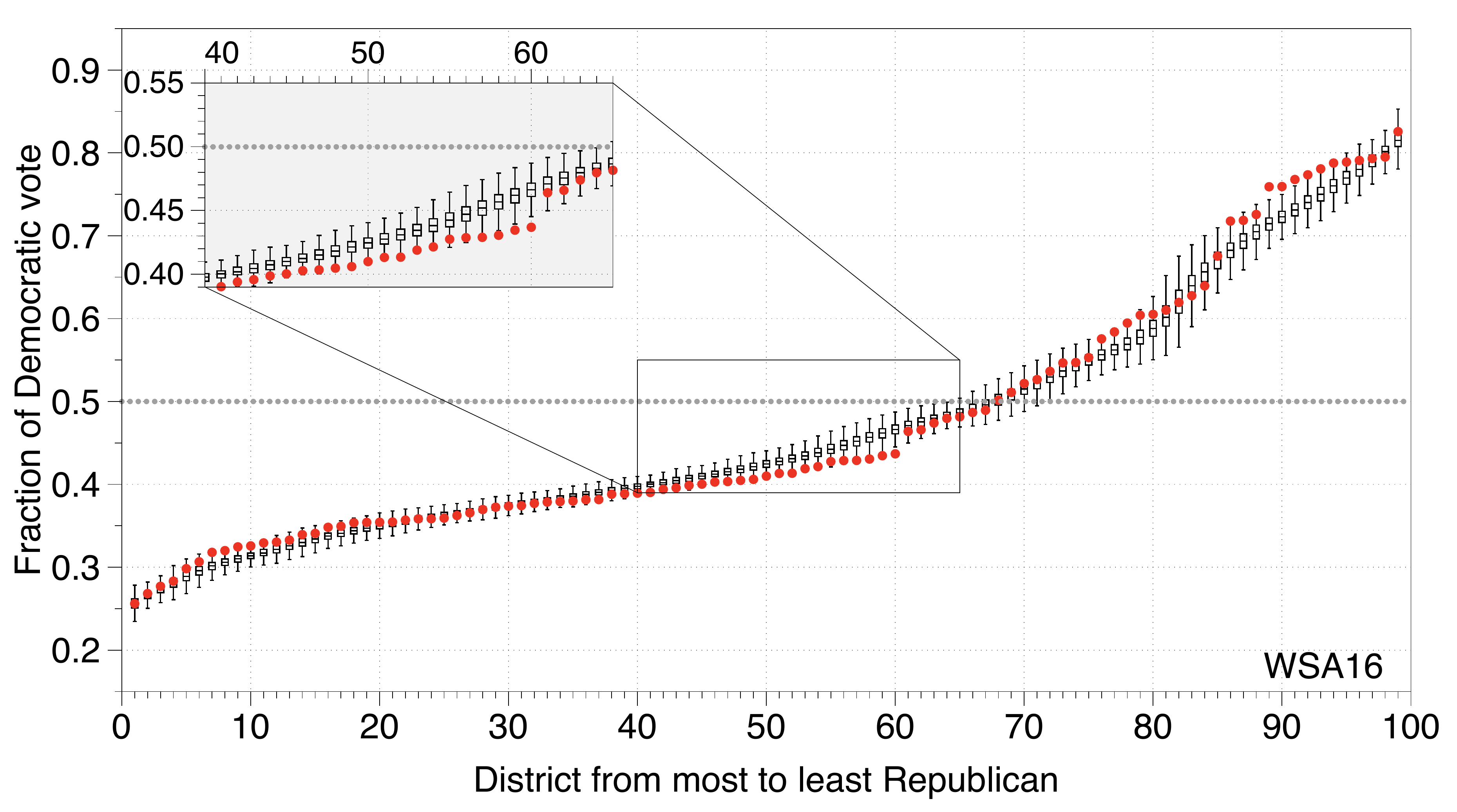}}
  \caption{Box-plot summary of districts ordered from most
         Republican to most Democratic, for the voting data from WSA12
         (top), WSA14 (middle), WSA16 (bottom). We compare our statistical results
         with Wisconsin redistricting in each case.}
\label{fig:boxPlots}
\end{figure}

The resulting
99 box-plots arranged on one graph for WSA12, WSA14, and
WSA16 give insight into the inherent geopolitical geometry of Wisconsin due to the
interaction of the state's geometry with population density and
partisan distribution (Figure~\ref{fig:boxPlots}). We see that
typically there are at least 25 districts with less then 40\%
Democratic vote. 

These plots give provide a method to determine the typical partisan makeup of 
 each district. This inherently reflects the geopolitical structure of
 the state. For a given redistricting plan, if a given district's
 percentage falls below the horizontal  50\%  line then the district
 elects a Republican. If it falls above the line, it elects a Democrat.
This plot has proven useful in detecting redistricting plans
 with packed or fractured districts. See \cite{QuantifyingGerrymandering,2017arXiv170403360B}. In some sense, they give
 quantitative definitions to these concepts.

If a given district's
Democratic vote percentage is at the bottom or below the box plot, the
district has fewer Democrats than expected. If the percentage
is above or at the top of the box plot, the district has more Democrats
than expected. 
It is clear that the Wisconsin
 Act 43 redistricting plan produces election results with Democratic votes
 depleted from the center of the plot and places those votes in the
 districts which already have a large number of Democratic voters. 

We also understand why the actual results of the WSA12 elections were not representative
while the actual results of the WSA14 and WSA16 elections were representative. It is simply a result
of where the 50\% line hits the box plot graph in
Figure~\ref{fig:boxPlots}. If the 50\% line crosses the graph in  the region in which the
location of the current Wisconsin redistricting plan (the red dots) falls outside of
the boxplot (which encodes typical behavior) then the results will be
anomalous. This is the case in WSA12 but not in WSA 14 and WSA16. On
the other hand, in all three years the district corresponding to the 50th
seat, which dictates who is in the majority, is always an outlier,
requiring less Republican votes than expected.

\subsection{Inherent Lack of Proportionality}
Notice that there is a structural tilt to the Republicans -- in all of our analyses,  a 50\% 
vote fraction for both parties leads to a majority of Republicans. 
We see that one only needs the Republican vote to be around
47\% to 49\% to obtain 50 seats with the structure of the WSA12 votes over the 
majority of redistricting plans. Similarly, WSA14 and WSA16 require between 46\% and 47\% 
Republican vote fractions and between 45\% and 46\% Republican vote fractions, 
respectively, to obtain 50 seats.  This shows clearly that it
is not reasonable to expect that 50\% of the vote leads to 50\% of the seats.
This does not explain all of the shift in the Republican favor
produced by the Wisconsin Act 43 redistricting plan. 
Our analysis allows us to separate out the effect of the geopolitical landscape, and to show that the Act 43 map generates extreme partisan asymmetry above and beyond this effect.

\subsection{Exploring Parity}
To further explore the impact of the structure in
Figures~\ref{fig:contShiftPlot} and~\ref{fig:boxPlots}, we explore
two ideas around parity. We begin by shifting the votes in WSA12,
WSA14, and WSA16 so that there are an equal number of redistricting plans in the
ensemble in which the Republicans and Democrats are in the majority.
When shifting the votes in this way, the Wisconsin Act 43
redistricting produces significantly more Republican seats -- 56 with
WSA12, 57 with WSA14, and 54 with WSA16. In the first two cases, this
is a result seen in very few redistricting plans of the ensemble redistricting plans while in WSA16
it has a very low probability (that is to say a relatively small  number redistricting
plans compared  to the just under 20,000 plans considered in the ensemble). 

Of course, one could perform a similar analysis around another point
than the 50\% mark. One can see whether if the Wisconsin redistricting would
still be an outlier when the votes are centered around a different
line by drawing a vertical line in
Figure~\ref{fig:contShiftPlot} at a different value and noting where
the thin black line corresponding to the Wisconsin Act 43 redistricting
crosses this vertical line. For instance for WSA12, any vertical line
up to at least 52\% and above 41\% results in a result with Wisconsin Act 43
which is well outside the results of 90\% of the ensemble --
very few redistricting plans which  give this result exist in the ensemble.
Similar in WSA14 from about 43\% to 50.5\% the results produced by Wisconsin
Act 43 are outliers as they lie outside the region containing 5\% to 95\% of the
ensemble. Lastly for WSA16, from 42\% to 50\% the Wisconsin redistricting
produces results which are outside the  region containing 5\% to 95\% of the
ensemble. In all cases the results are skewed to the Republicans
precisely in the region where the Democrats threaten to move into the majority.

\begin{figure}[ht]
  \centering 
  \includegraphics[width=.65\linewidth]{\dir{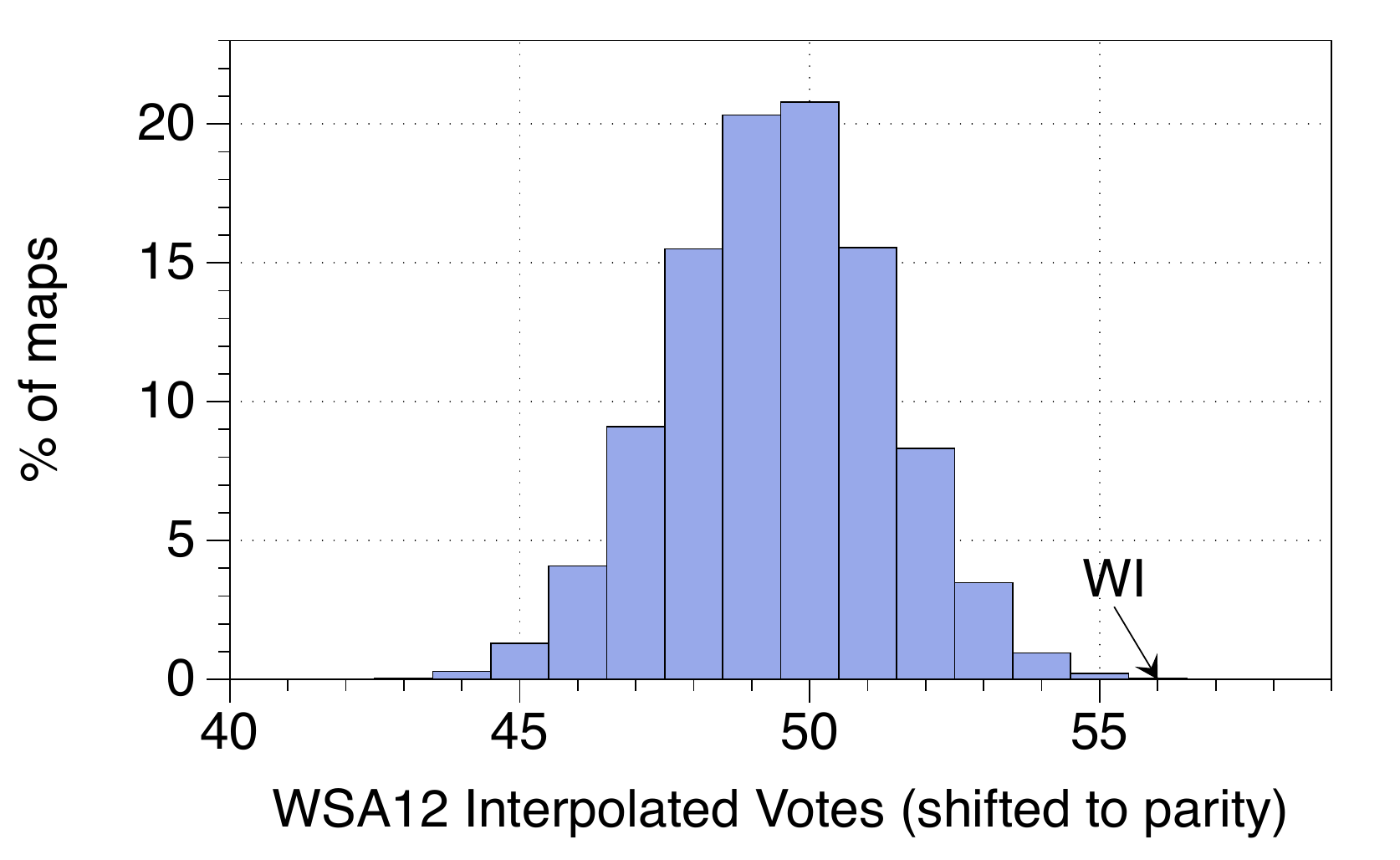}}
  \includegraphics[width=.65\linewidth]{\dir{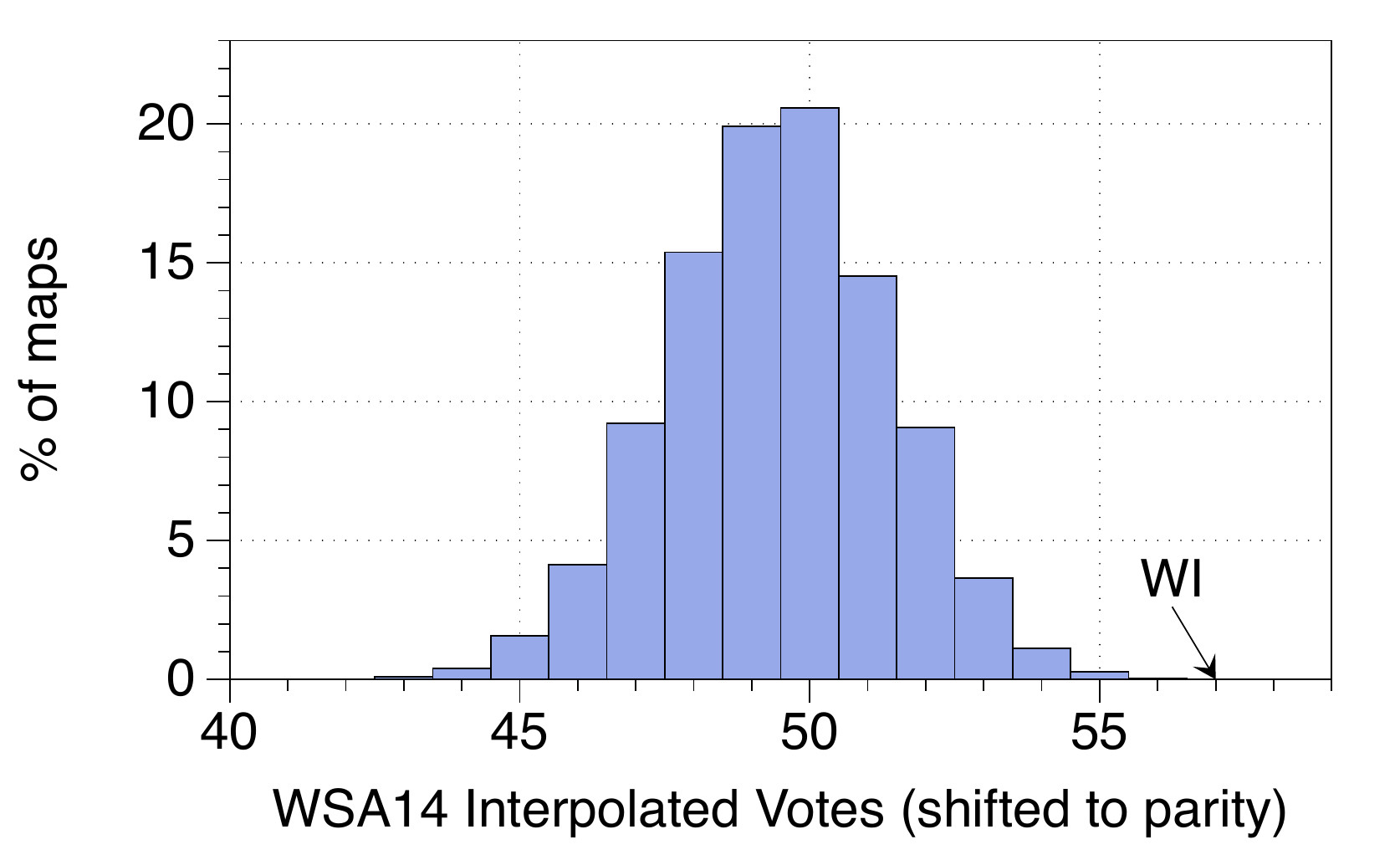}}
  \includegraphics[width=.65\linewidth]{\dir{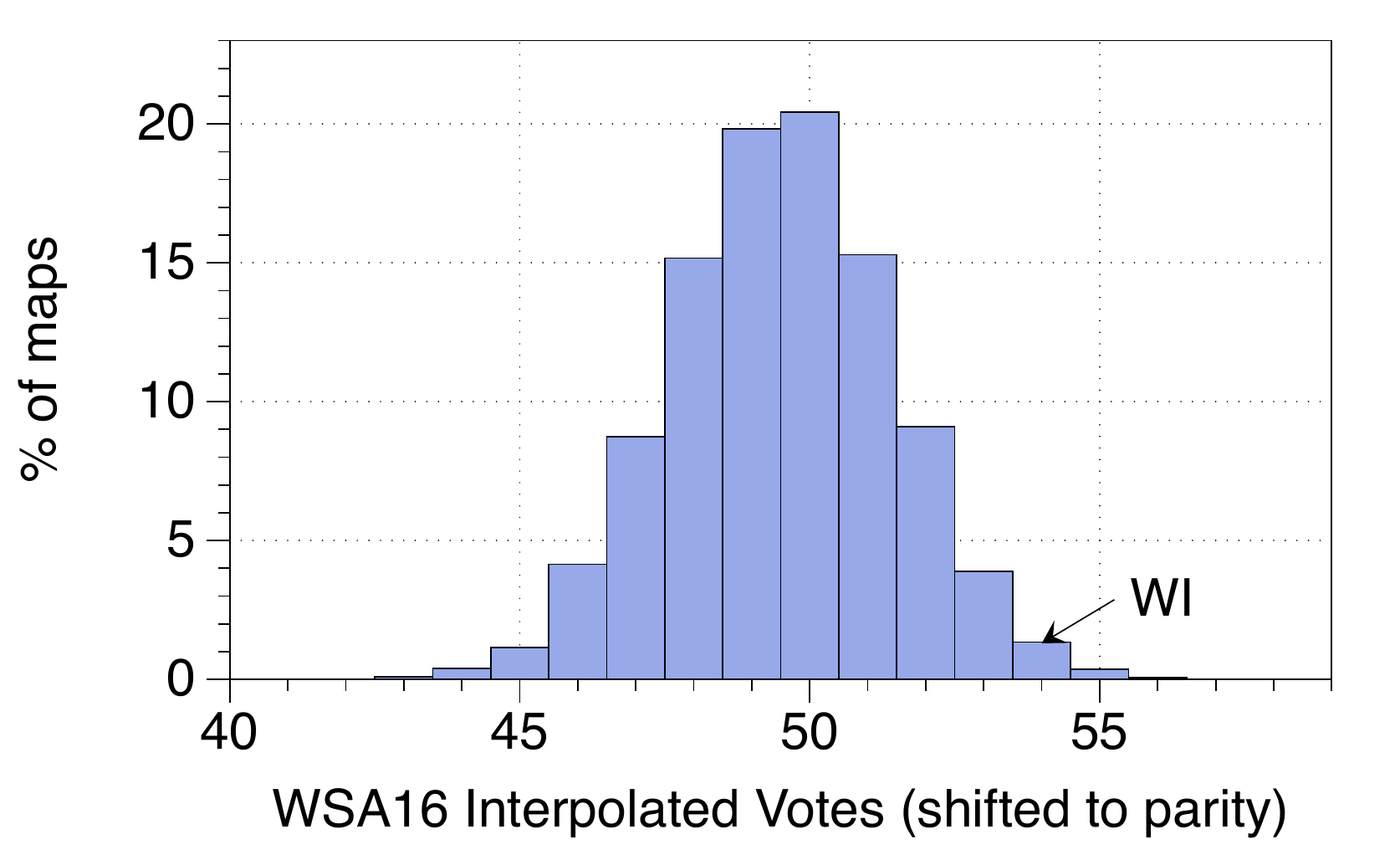}}
  \caption{Histogram of Republican seats won when WSA12 (top), WSA14 
    (middle), WSA16 (botom) shifted so that half of the redistricting plans lead to a majority for either party.}
  \label{fig:histeven}
\end{figure}

\begin{figure}[ht]
  \centering 
  \includegraphics[width=.65\linewidth]{\dir{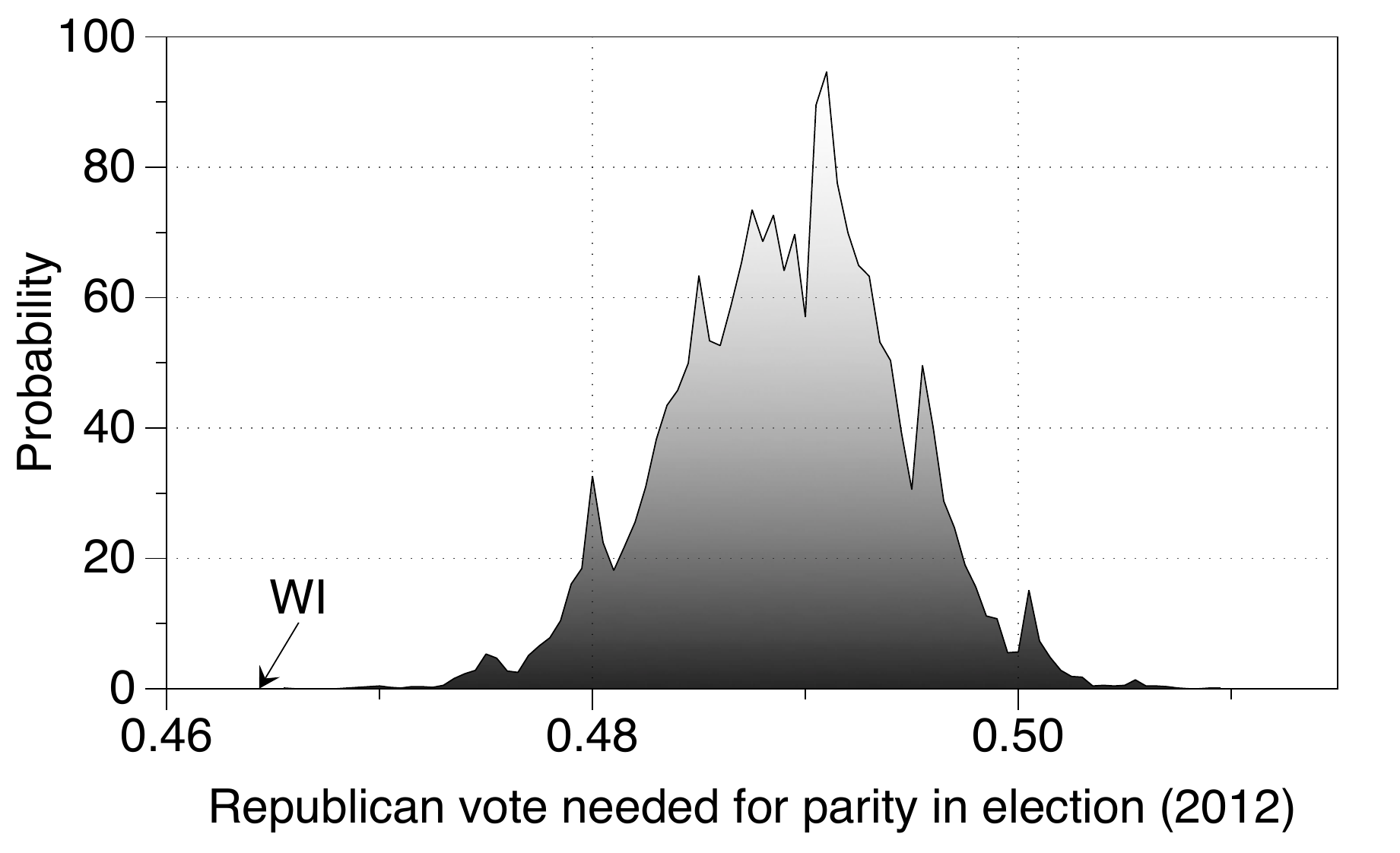}}
  \includegraphics[width=.65\linewidth]{\dir{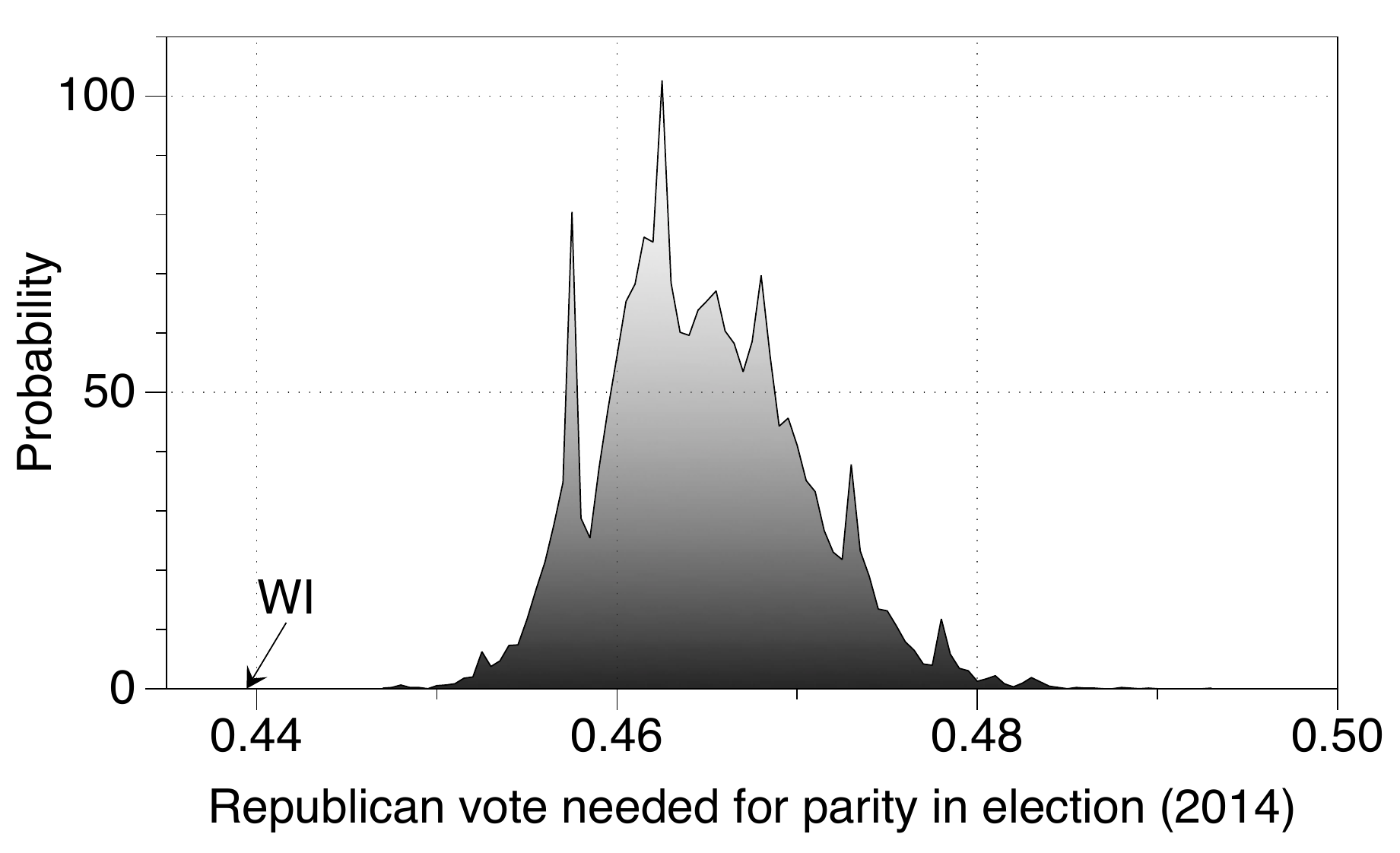}}
  \includegraphics[width=.65\linewidth]{\dir{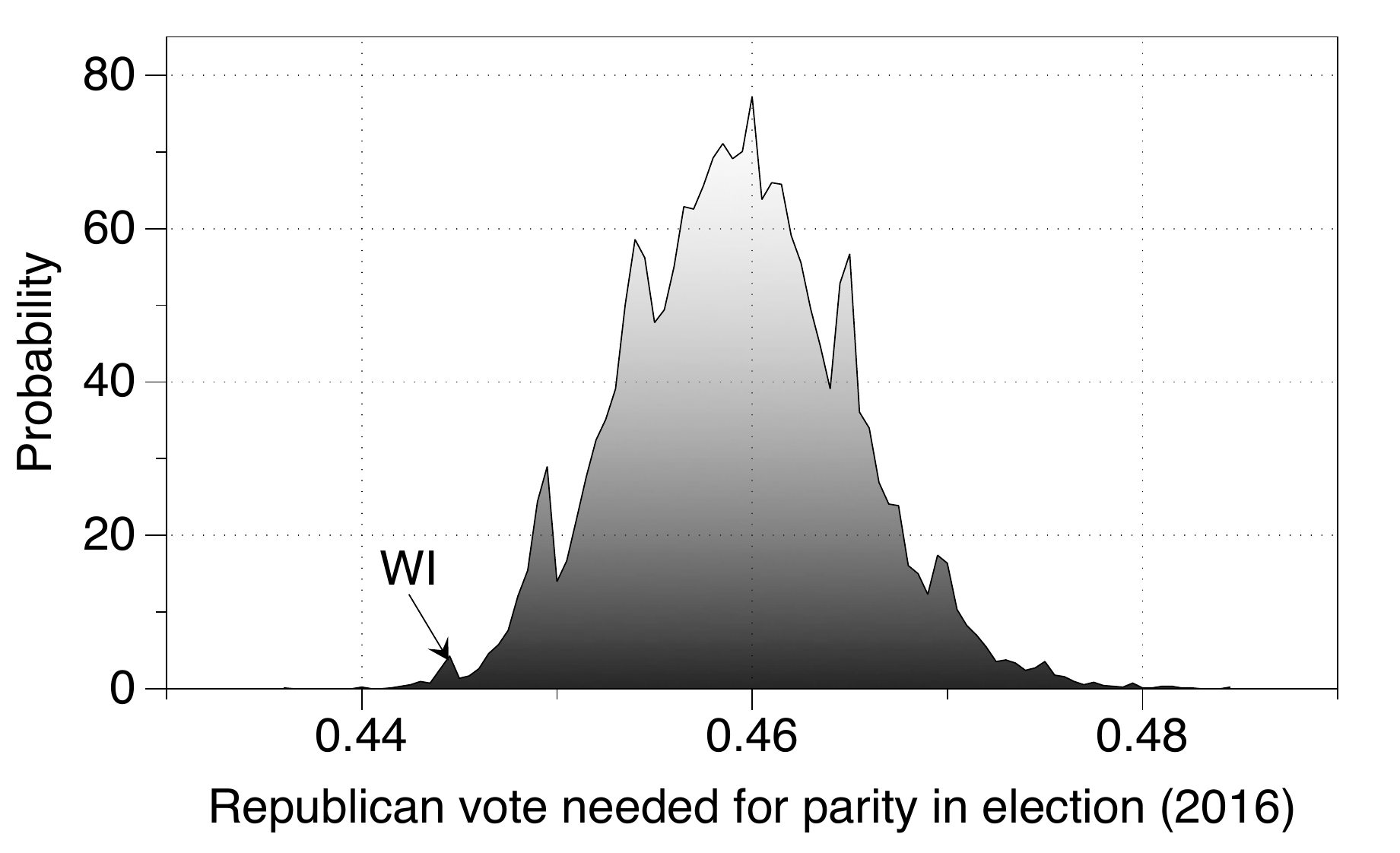}}
  \caption{Votes fraction needed so both parties have an equal chance at majority}
  \label{fig:histevenShift}
\end{figure}

A complimentary perspective is instead to ask to what percentage
of Republican vote does one have to shift the election to produce a
50/50 split of the seats with a given redistricting. A histogram of
the quantity tabulated over the ensemble is shown in
Figure~\ref{fig:histevenShift} along with the percentage needed for the
Wisconsin Act 43 redistricting plan. Again we see there is a systematic tilt
towards the Republicans built into the geopolitical structure of the
state. However, in all cases the percentage needed to produce
parity in the Wisconsin Act 43 plan is abnormally low.

\section{Summary Statistics}
\label{sec:sumary-statistics}
We now develop a number of summary statistics that 
highlight and make quantitative  the graphical analysis developed in the last section. 

\subsubsection*{Gerrymandering Index} We begin by calculating the Gerrymandering Index developing in
\cite{2017arXiv170403360B}. It measures the extent to which a
particular redistricting has districts whose vote margins for each election deviate from
what is expected in Figure~\ref{fig:boxPlots}. For a
given election, the square of the Gerrymandering index is the sum of
the square deviations of each of the sorted Democratic percentages
from the means of the marginals in the ninety-nine box-plots in Figure~\ref{fig:boxPlots}.
To contextualize the Gerrymandering Index, we situate a given score
within the distribution of scores from our ensemble of redistricting plans. Redistricting plans
which have unusually large Gerrymandering index should be view as Gerrymandered. 
The percentage of the ensemble with Gerrymandering Index worst than
the Wisconsin Act 43 redistricting is shown in Table~\ref{tab:index} for a
number of different sets of votes from different years. In all cases,
the  Wisconsin Act 43 redistricting is seen to have an unusually high level of
the Gerrymandering Index. 

\begin{table}[h]
  \centering 
  \begin{tabular}{lp{2cm} p{2cm} p{1.5cm}}
Voting data & \% more gerrymandered & \%  less representative & Rep. Vote Fraction.\\
    \midrule 
    WSA16 & 0.01 & 87.98 & 52.91\\
    GOV12 & 0 & 48.61 & 52.59 \\
    USH14 & 0.09 & 46.78 & 51.89\\
    WSA14 & 0 & 82.34 & 51.28\\
USS16 & 0 & 40.93 & 51.09\\
WSA12 & 0 & 0.44 & 50.05 \\
PRE16 & 1.52 & 60.00 & 49.91\\
USH12 & 0 & 25.02 & 48.68\\
USS12 & 0 & 0 & 46.65\\
PRE12 & 0 & 0 & 45.98\\
\bottomrule \\
\end{tabular}
 \caption{We show the percentage of redistricting plans within the ensemble
   that are (i) more gerrymandered and (ii) less representative than Wisconsin
   Act 43 redistricting; we also display the republican vote fraction.  According to all vote counts, the current Wisconsin plan is highly gerrymandered.  There is a strong correlation between Republican vote fraction and Representativeness. (PRE =
   President, WAG=Attorney General, GOV= Governor, USS= US Senate,
   USH=US House)}
\label{tab:index}
\end{table}

\subsubsection*{Representative Index}
The
Gerrymandering Index directly measures  how anomalous the partisan
composition of a redistricting is. It is possible for a redistricting
to be gerrymandered, yet still be
representative of the vote count, as we have seen in Figure~\ref{fig:hist} for WSA14 and WSA16.
Therefore, we also measure how representative a redistricting is in the context of different vote counts.

In \cite{2017arXiv170403360B}, we also define a Representative Index
which quantifies how representative the result obtained by using a
particular redistricting and vote combination is. It is essentially the
distance from the mean value in the histograms in
Figure~\ref{fig:hist} when one extends the number of seats won by a
given party to a continuous variable in a natural way. See
\cite{2017arXiv170403360B} for the details. As with the Gerrymandering
Index, in Table~\ref{tab:index} we postion the Representative Index
inside the ensemble of redistricting plans by reporting the percentage of
the ensemble with a larger index. It is worth noting that
Table~\ref{tab:index} shows the same dependence of representativeness
reflected in the histograms in Figures~\ref{fig:allRealData} and
\ref{fig:shifHist} and the plots in Figure~\ref{fig:contShiftPlot}.
As the global percentage of Republicans decreases towards 50\% the
representative score begins to drop.  The effect is not strictly
monotone as the geopolitical structure of each vote also plays a role.
 
\subsubsection*{Representativeness Measured Across Shifts} 
From the preceding section, it is clear that the  overall percentage
of the vote as well as its geopolitical
structure can have a large effect on the perceived representativeness of
a redistricting, even when the Gerrymandering Index reports a high level of
gerrymandering. To control for this, we consider shifts of a
given collection of votes much in the spirit of
Figure~\ref{fig:shifHist}. 

Rather than use the Representative Index from
\cite{2017arXiv170403360B}, we consider an alternative
formulation which measures the negative log probability of the observed elevation
outcome using the probabilities from our ensemble. We then sum
these values over a number of shifts of the original election.  The logarithmic measure  more
naturally lives on the same scale across different elections and hence
seems more appropriate for this context. This measure, which we will
denote by $H$, is essentially
an average log-likelihood across the different shifts 

We  compliment this nonpartisan statistic with one designed to measure
deviations in the Republicans's advantage, denoted by $L_\rep$, and one to measure deviations in the Democrats's
advantage, denoted by $L_\dem$. In
Table~\ref{tab:index}, we see based on the $H$
statistics that the Wisconsin Act 43 districts are outliers being much less
representative than most of the redistricting plans in the ensemble. The
$L$ statistic shows that the Wisconsin redistricting is
tilted to favor the Republicans. In one year the $L_\dem$ raises to
the almost 75\%; however the box-plots in Figure~\ref{fig:boxPlots}
show that the benefit to Democrats comes in elections where the
Republicans hold a strong majority in any event, so that the benefit
does not affect majority control.

To capture the representativeness over a range of election outcomes,
we  consider shifted election votes over a range of outcomes. We
 consider a measure which registers both the worst-case deviation
from the typical and one which measures the average deviation.

%shorten sentence
Fixing a set of votes to evaluate election outcomes, we define the
index $\ell_{\rep}$ to be the minimum, overall shifts of the percentage
Republican vote between 45\% and 55\%, of the probability that the number of
Republican seats for a given map is greater than one drawn from our
distribution. We estimate this probability using the ensemble we
generated.  We then define $L_{\rep}$ to be the fraction of maps in our
ensemble for which $\ell_{\rep}$ is greater than it is for the map in
question. We define $\ell_{\dem}$ and $L_{\dem}$ in the same fashion but
with Republicans replaced by Democrats.  

The $L$  statistics described above compare the
worst case between two redistricting plans and are inherently one-sided, hence
the Democratic and Republican versions. It is also useful to consider
a statistic which is an average  over a range of shifts. Again fixing
a set of votes and a redistricting to be investigated, we define $h$ to be the
sum over a set  of shifts of the logarithm  of the probability that
two of the redistricting plans in question produce the same number of
Republicans as a random redistricting drawn from our distribution. As
with the preceding statistic, we determine a sense of scale for $h$ by
defining $H$ to be the probability that the $h$ of a given
redistricting is greater than a randomly drawn redistricting from our
ensemble.

\begin{table}[h]
\centering
\begin{tabular}{l | lll}
 & $H$ & $L_\rep$ & $L_\dem$\\
 \hline 
WSA12 & 100\%& 99.833\% & 3.253\%\\
WSA14 & 99.990\% & 99.765\% & 72.785\% \\
WSA16 &99.223\%&  99.656\% & 7.892\% \\
\end{tabular}
\vspace{1em}
\caption{Summary statistics measuring representativeness for Wisconsin Act 43
redistricting. These numbers give the percent of redistricting plans the Wisconsin Act 43 redistricting is worse than in terms of average ($H$), Republican favoritism ($L_{rep})$, and Democratic favoritism ($L_{dem}$).}
\label{tab:sumstat}
\end{table}

The results in Table~\ref{tab:sumstat} show that the results are
clearly anomalous. The values of $H$ for the Wisconsin plan are extreme
outliers within our ensemble. By detecting unrepresentativeness over
a range of shifts, the $H$ statistic assess the level of
gerrymandering in the range of total vote fractions where elections
typically occur. 

We now clarify these definitions by restating them in more
mathematical notation. We begin by fixing some notation.
For any redistricting $\pi$ and $s \in \R$, we let
$\pi+s$ to be the vote obtained by shifting the partisan vote $s\%$ to
the Republicans. Let $\rep(\pi)$ and $\dem(\pi)$ denote respectively the
total percent Republican or Democratic vote
in the election $\pi$. Now for any redistricting $\xi$, we let
$\mathrm{Rep}(\xi,\pi)$  and  $\mathrm{Dem}(\xi,\pi)$ be the total
number of seats won by respectively the
Republicans and Democrats with vote $\pi$ and redistricting $\xi$. 

Now we define 
\begin{align*}
  \ell_{\rep}(\xi,\pi) &= \min_{s \in [45,55]-r(\pi)}  \pr \Big( \mathrm{Rep}\big(\Xi,\pi+s\big) 
  \geq \mathrm{Rep}(\xi,\pi+s\big) \Big)\\
\ell_{\dem}(\xi,\pi) &= \min_{s \in [45,55]-r(\pi)} \pr \Big( \mathrm{Dem}\big(\Xi,\pi+s\big) 
  \geq \mathrm{Dem}\big(\xi,\pi+s\big) \Big)
\end{align*}
where $\Xi$ is a redistricting chosen uniformly from our ensemble and
$[45,55]-r(\pi)$ is compact notation for the set of shifts $[45 -r(\pi),55 -r(\pi)]$.
We
then situate these probabilities in the ensemble by defining 
\begin{align*}
  L_{\rep}(\xi,\pi)&= \pr\big( \ell_{\rep}(\Xi,\pi)  \leq
                    \ell_{\rep}(\xi,\pi)\big) \\
  L_{\dem}(\xi,\pi)&= \pr\big( \ell_{\dem}(\Xi,\pi)  \leq  \ell_{\dem}(\xi,\pi)\big) 
\end{align*}
where $\Xi$ is again a randomly chosen redistricting from the
ensemble.

To define the averaged representative index, we define the average
log-likelihood 
\begin{multline*}
  h(\xi,\pi) =- \frac1{|I|}\sum_{s \in I-r(\pi)} \log\, \pr\Big( \mathrm{Rep}\big(\Xi,\pi+s\big) = \mathrm{Rep}\big(\xi,\pi+s\big) \Big)
\end{multline*}
where $\Xi$ is chosen according to our distribution on redistricting
plans and $I=\{45,45.5,\dots,54.5,55\}$, $I-x$ is the shifted set
defined as before by $I-x=\{ y-x: y \in I\}$, and $|I|$ is the number
of points in $I$.
 We then situate these in the ensemble by defining
\begin{align*}
  H(\xi,\pi) &= \pr\big( h(\xi,\pi) \geq  h(\Xi,\pi)\big)\,.
\end{align*}
In calculating $H$, we extrapolate  the observed histogram using
a Gaussian tail approximation whenever a values is needed outside the
range observed in the histogram.

We report the summary statistics in Table~\ref{tab:sumstat}.  We find
that the Wisconsin Act 43 redistricting is an extreme outlier in terms
of how probable it is to observe its value of $H$.  We also find that in the
worst case, it can benefit the Republicans by more than 99\% of all
redistricting plans in our ensemble.  Conversely, when shifting
between 45\%-55\% of the vote fraction, the Democrats are
significantly impeded in WSA12 and WSA16, and are aided to a much
lesser degree in other elections and vote shifts.  We remark that when
we re-examine Figure~\ref{fig:contShiftPlot}, the Democrats are only
`aided,' once the Republicans have obtained a super majority, as can
be seen by the thin continuous line falling below the 90\% region.

\section{Generating the Ensemble of Redistricting Plans}\label{sec:generatingEnsemble}
Our method begins by first placing a probability distribution on 
all the reasonable redistricting plans. The probability distribution will be
concentrated on redistricting plans which better satisfy the design
certain specified in the laws and legal precedents covering
redistricting plans in Wisconsin.
We then draw an ensemble of redistricting using the classical Markov
Chain Monte Carlo algorithm. The frequency of redistricting plans with
different qualities will depend on how well those districts satisfy
the design criteria.
\subsection{The Distribution on Redistricting Plans}
\label{sec:distr-redistr}
Following the prescription from \cite{2017arXiv170403360B} (see also
\cite{MattinglyVaughn2014,QuantifyingGerrymandering}), we consider
distributions with a density proportional to 
\begin{align*}
  e^{-\beta J(\xi)}
\end{align*}
where $\xi$ is the function which assigns to each ward a district
which we label with the numbers 1 to 99 for convenience. The score
function $J$ will be the sum of a number of different score
functions
\begin{align*}
  J(\xi) = &w_\mathrm{comp}J_\mathrm{comp}(\xi) +w_\mathrm{pop} J_\mathrm{pop}(\xi)\\&+  w_{county}J_\mathrm{county}(\xi)
  +w_\mathrm{vra}J_{\mathrm{vra}}(\xi) 
\end{align*}
where $J_\mathrm{comp}(\xi)$ measures compactness, $J_\mathrm{pop}(\xi)$ measure
population deviation from the ideal, $J_\mathrm{county}(\xi)$ the number of
counties split across counties, and $J_\mathrm{vra}(\xi)$ measures the
compliance with the Voting Rights Act (VRA); the $w_i$'s are positive weights. In all cases, low scores
will correspond to better compliance with the associated design criteria.

We will use the population and compactness score functions
from  \cite{2017arXiv170403360B}. The county and VRA score functions
will versions of those from
\cite{2017arXiv170403360B} with modifications to adapt to the details
of the Wisconsin redistricting context. The Wisconsin State Assembly (WSA)
districting required that many counties and towns be split into more
than two districts (in contrast to the work in
\cite{2017arXiv170403360B}). Hence minor alterations were required to
our previous score functions.

We determine the weight parameters with a nearly identical process to that described in \cite{2017arXiv170403360B}.   We have determined $w_{pop}=2200$, $w_{comp}=0.8$, $w_{county} = 0.6$, $w_{VRA}=100$.

\subsection{Markov Chain Monte Carlo Sampling}

We sample redistricting plans according to the algorithm presented
\cite{2017arXiv170403360B}. For simulated annealing parameters, we take 20,000 accepted steps 
at $\beta=0$, 80,000 accepted steps as $\beta$ linearly 
increasing to one, and $20,000$ steps for $\beta=1$ ( see 
\cite{2017arXiv170403360B}  for more details about the meaning of 
these parameters).  In our reported ensemble we take 19,184
samples.  In Section~\ref{sec:robustness-results}, we show evidence
that this is sufficient to correctly sample the distribution on redistricting plans.

\subsection{Redistricting Plans in the Ensemble}
\label{sec:constr-ensemble}

We ensure that the districts are contiguous, all redistricting plans are
more compact than the Wisconsin Act 43 plan.  We only kept samples
such that the maximum population deviation is below 5\%.  To account
for the VRA, we only retain redistricting plans containing six districts that have at least 40\% African Americans and one district that has at least a 40\% Hispanic population.  All sampled redistricting plans are described at the ward level, and no ward is split.  

With the above criteria, we have account for all districting criteria present in the Wisconsin constitution, with the exception of splitting townships.  We also gather a smaller number of samples (2043) from a distribution that concentrates on redistricting plans that also preserve townships.  The township consideration requires an additional term in the score function, and is similar in form to the county splitting energy.   We compare the effect of preserving townships below in Section~\ref{sec:robustness-results}.

\section{Interpolating Election Data}
\label{sec:interp}
We now explain how  we interpolate the election data which is missing due
to unopposed races. Let $V_{\tt}(i)$, $V_{\dd}(i)$, and $V_{\rr}(i)$ be 
respectively the total vote, the Democratic vote, and the Republican vote in
ward $i$. We split the ward indices into the good $\mathcal{G}$ and
$\mathcal{B}$ wards. Typically the wards in $\mathcal{B}$ are the wards where
the race is unopposed. We also make use of a second set of
voting data $(U_{\tt}(i), U_{\dd}(i), U_{\rr}(i))$ for which no data
is missing. We begin by considering the  pairs ${(U_{\tt}(i),
  V_{\tt}(i)) : i \in \{1,\dots 99}\}$ which we assume to be sorted
by the first value. To interpolate $V_{\tt}(i)$ for some $i \in
\mathcal{B}$, we select the pairs 
\begin{multline*}
 \big\{ (U_{\tt}(i_{-2}),  V_{\tt}(i_{-2})), (U_{\tt}(i_{-1}),
  V_{\tt}(i_{-1})), \\
 (U_{\tt}(i_{1}),  V_{\tt}(i_{1})), (U_{\tt}(i_{2}),  V_{\tt}(i_{2}))\big\}
\end{multline*}
where $i_1$ and $i_2$ are the next two elements in the increasing
ordered sequence of $U_{\tt}$ values after $U_{\tt}(i)$ so that
$i_1, i_2 \in \mathcal{G}$. Similarly $U_{\tt}(i_{-2})$ and
$U_{\tt}(i_{-1})$ are the previous two elements in the ordered
sequence again so that both indices are in $\mathcal{G}$. If no such
point exists, we proceed with the points we have. We then
perform a linear least-squares fit to this collection of
points. Observe that there are always at least two points in the
collection. We then evaluate this linear fit at the point
$U_{\tt}(i)$ to obtain our estimate of $V_{\tt}(i)$ which we will denote
by $\widehat V_{\tt}(i)$. Then, in the same fashion, we estimate
$\rho_{\rr}(i) =U_{\rr}(i)/U_{\tt}(i)$ and
$\rho_{\dd}(i) =U_{\dd}(i)/U_{\tt}(i)$ with
$r_{ \rr}(i) =V_{\rr}(i)/V_{\tt}(i)$ and
$r_{\dd}(i) =V_{\dd}(i)/V_{\tt}(i)$ to obtain $\widehat \rho_{\rr}(i)$
and $\widehat\rho_{\dd}(i)$.  We then set $\widehat V_{\rr}(i)$ to be
the average of
$\mathrm{floor}(\widehat \rho_{\rr}(i)\widehat V_{\tt}(i))$ and
$\widehat V_{\tt}(i)) - \widehat \rho_{\dd}(i)\widehat V_{\tt}(i))$
and similarly $\widehat V_{\dd}(i)$ to be the average of
$\widehat \rho_{\dd}(i)\widehat V_{\tt}(i)$ and
$\widehat V_{\tt}(i)) - \widehat \rho_{\rr}(i)\widehat
V_{\tt}(i)$. For each choice of reference vote
$(U_{\tt}(i), U_{\dd}(i), U_{\rr}(i))$, we obtain such an estimate. In
some cases, we obtain multiple such estimates associated with different
reference votes. We then average all of the estimates to obtain a
finial estimate which we then round to the nearest integer.

To decide which of the many possible combinations of reference votes
$(U_{\tt}(i), U_{\dd}(i), U_{\rr}(i))$ produce the best results, we
also predict the values for the $i \in \mathcal{G}$ and select the
collection of reference votes which produces the smallest total
squared error. This leads to the choices of reference votes presented in Table~\ref{tab:interp}
in the following elections with unopposed elections.

\begin{table}
  \centering
  \begin{tabular}{l|l}
    Election to interpolate & Reference elections\\
    \midrule 
WSA12 & PRE12, USS12, USH12  \\
WSA14 & GOV14, WAG14 \\
WSA16 & PRE16, USS16\\
  \end{tabular}
\vspace{1em}
\caption{Data used to interpolate Wisconsin State Assembly date. See
  Table~\ref{tab:index} for abbreviations.}
 \label{tab:interp}
\end{table}

In 2012 and 2014, the interpolated votes yield the same number of
seats with the Wisconsin Act 43 maps as the original vote counts with the
unopposed races not interpolated. In 2016, the number of seats changed
from 64 to 67. To understand why this occurred, observe that  districts 54, 73 and 74 were uncontested by Republicans and thus went to the Democratic candidate; however the votes in these districts leaned Republican for the President and the Senate, explaining why the interpolated result disagrees with the actual result.

\section{Robustness of Results}
\label{sec:robustness-results}

\begin{figure}
\centering
\includegraphics[width=\figS\linewidth]{\dir{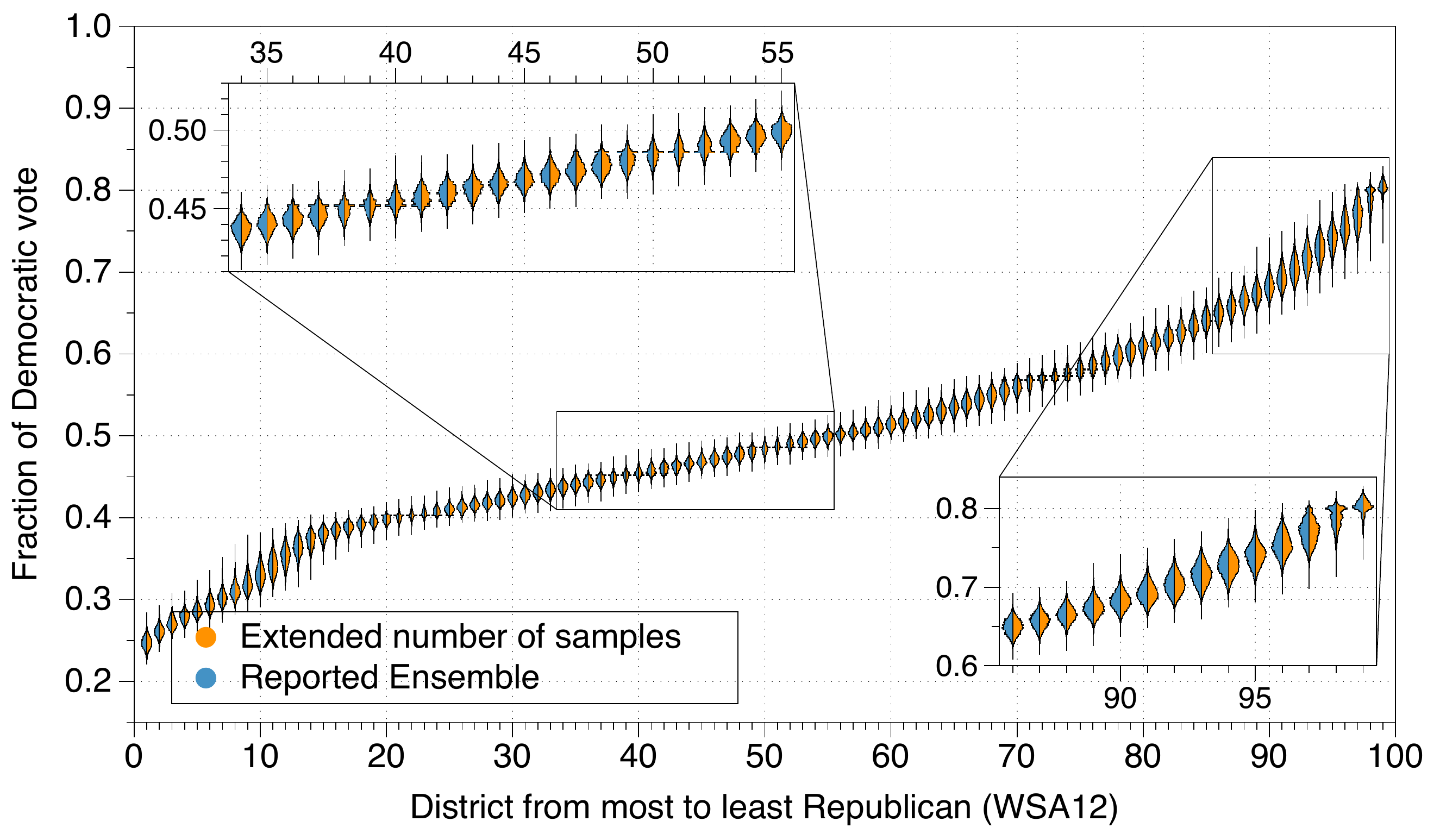}}
\includegraphics[width=\figS\linewidth]{\dir{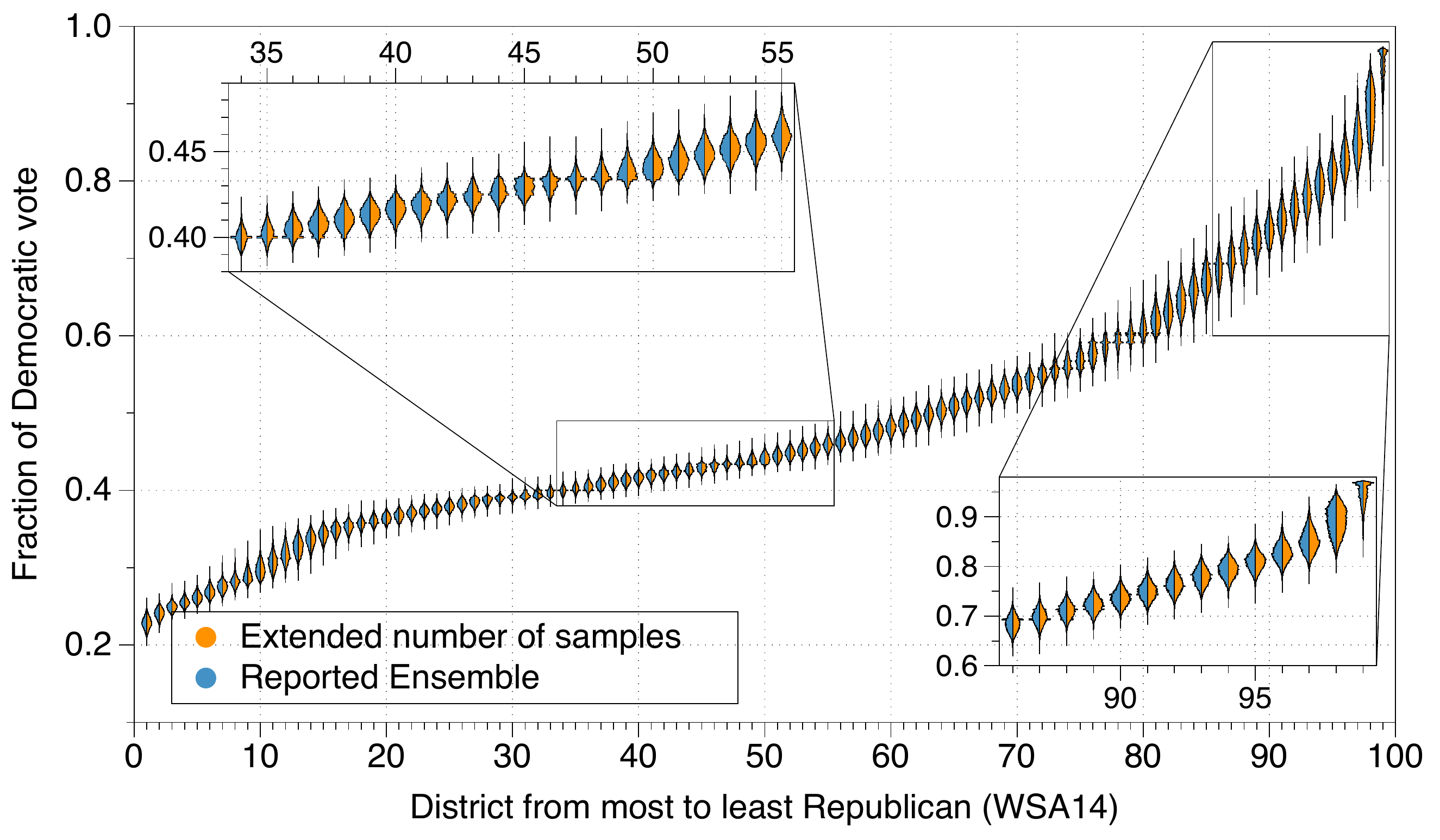}}
\includegraphics[width=\figS\linewidth]{\dir{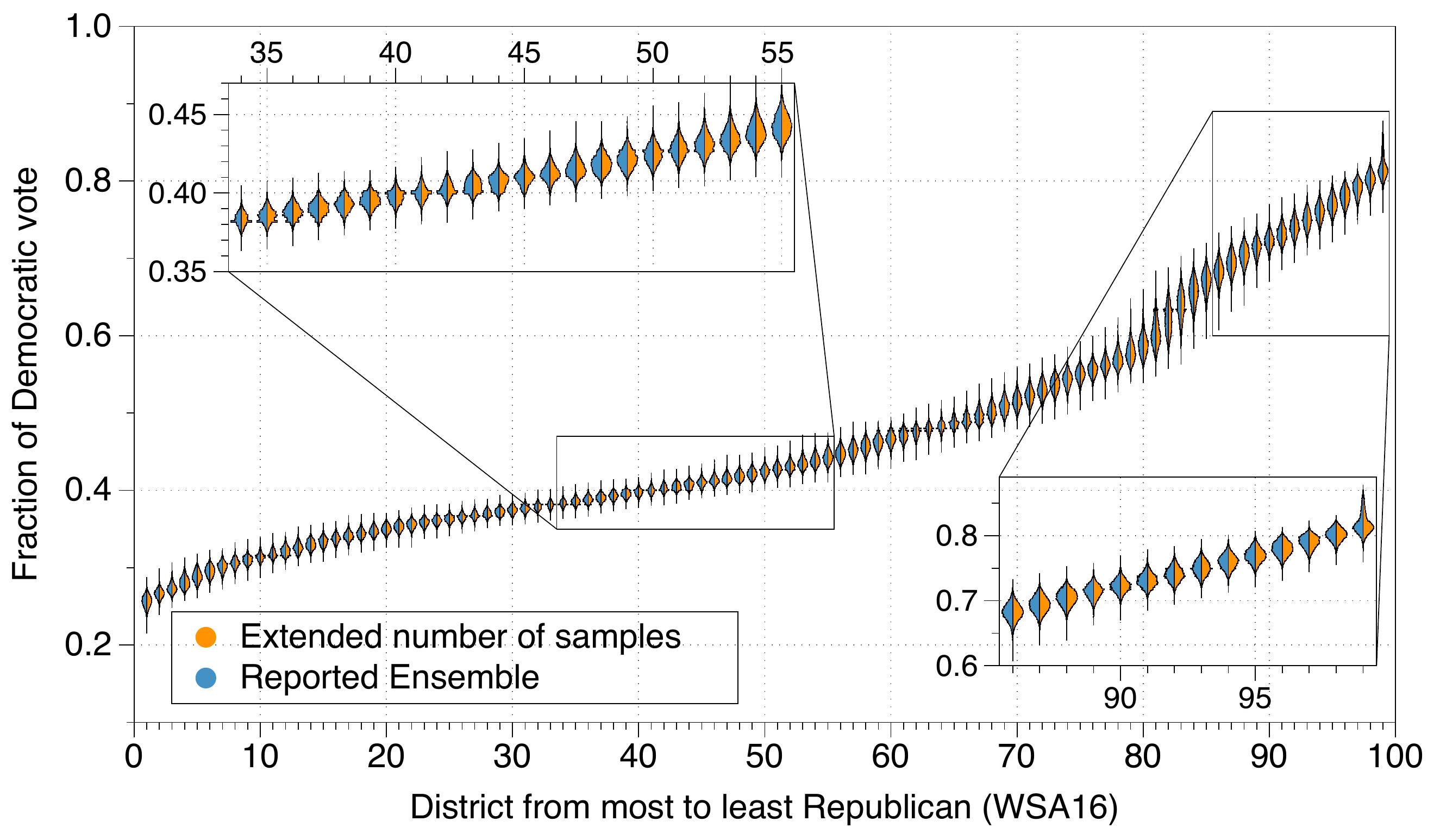}}
\caption{Testing the effect of using an ensemble with more samples.}
\label{extendedcomp}
\end{figure}

To check the robustness of our results, we (i) take longer runs, and (ii) generate a second ensemble in which we additionally account for town splitting.  In considering more runs, we extend our sampling algorithms to examine 84500 redistricting plans.  This extended sampling tests whether we have appropriately sampled the space of redistricting plans.  The box plots are the most detailed of our results and all other results may be derived from the data contained within them; to show even more detail we present marginal histogram plots.  We plot the marginal histograms of the extended samples compared with the reported samples in Figure~\ref{extendedcomp}.  We find the histogram structures are visually identical for WSA12, WSA14, and WSA16 voting data which provides evidence that we have appropriately sampled the space of redistricting plans.

To consider the effect of keeping townships contiguous, we add a fifth
term to the score function that is similar to the county splitting
score reported in \cite{2017arXiv170403360B}.  We consider townships
to be all wards with the same name within the shapefile provided by
the Legislative Technology Services Bureau \cite{WIShapeFileData}. For
example, using this criteria the city of Wausau is comprised of 41 wards.  The new score function is weighted with a value of 0.005, which we have found only marginally affects the overall districting compactness and keeps townships together in a similar way to that of the current plan in Wisconsin.  We sample 2043 redistricting plans that preserve townships.  

\begin{figure}
\centering
\includegraphics[width=\figS\linewidth]{\dir{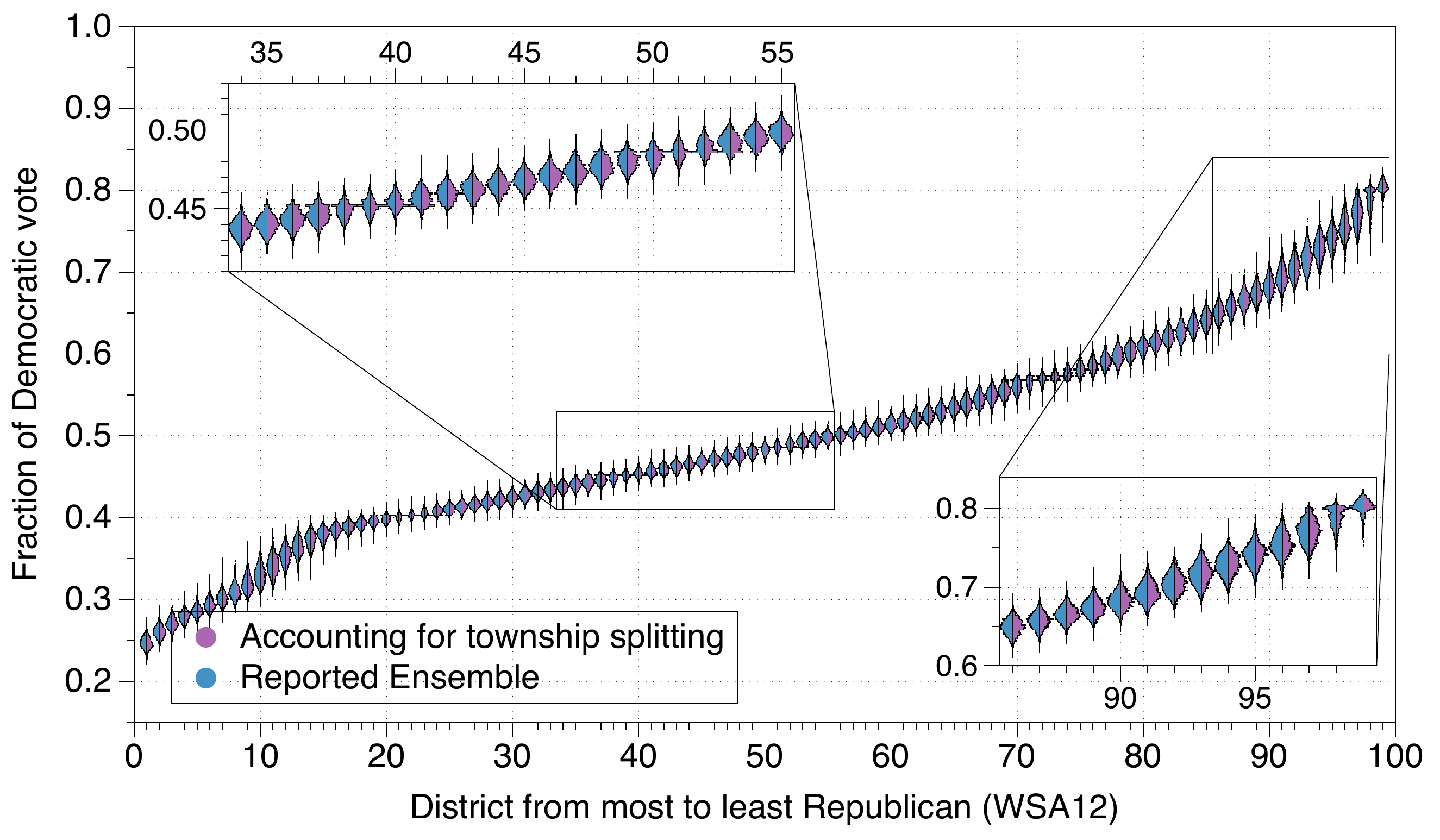}}
\includegraphics[width=\figS\linewidth]{\dir{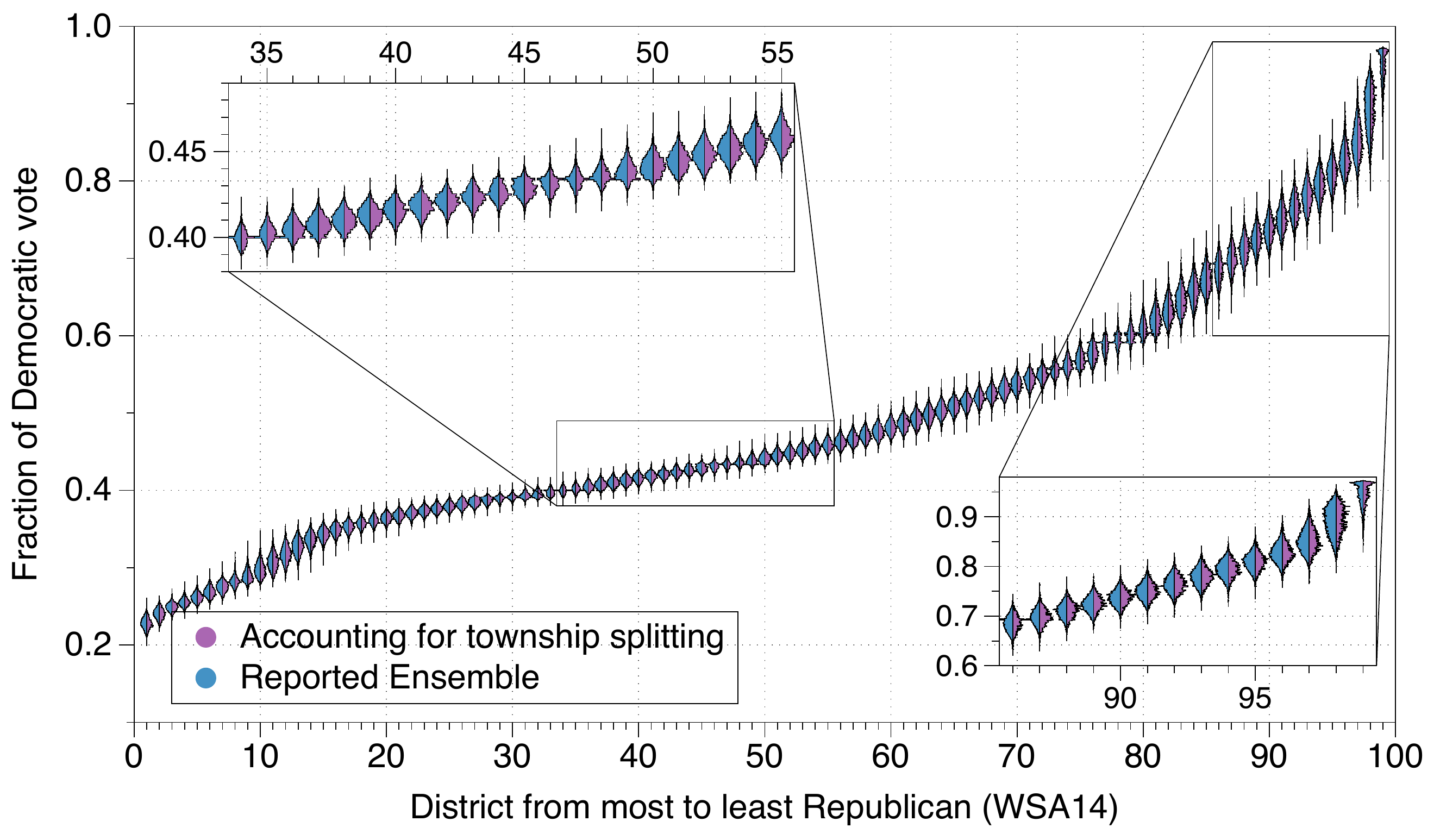}}
\includegraphics[width=\figS\linewidth]{\dir{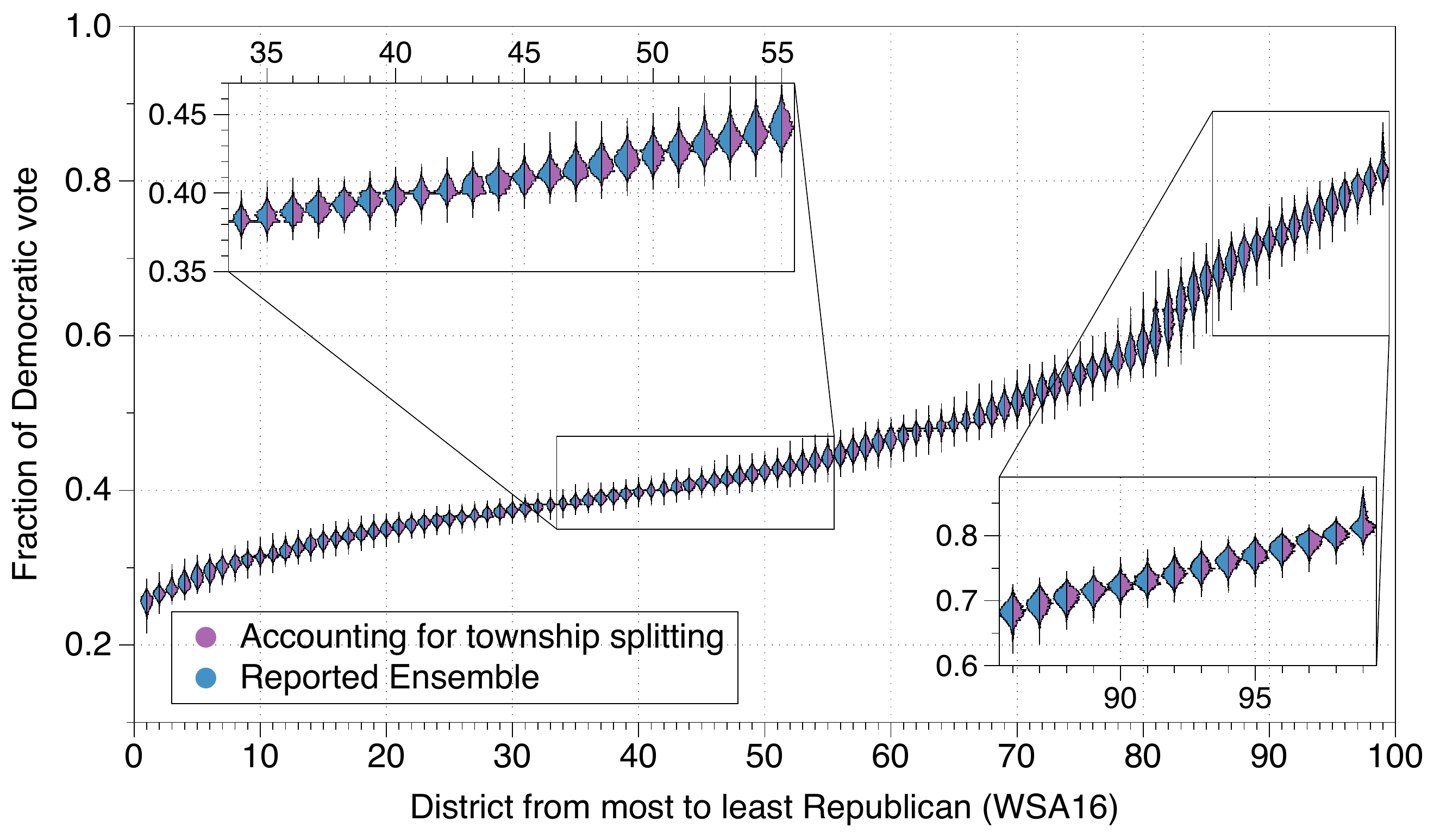}}
\caption{Testing the effect of favoring townships not being split by
  district boundaries on the districting results.}
\label{townsvsnotowns}
\end{figure}

We compare the marginal histogram plots when considering township splitting and for the ensemble we have reported above in Figure \ref{townsvsnotowns}.  We find the histogram structures are visually identical for WSA12, WSA14, and WSA16 voting data.  Because this new ensemble predicts identical district level results, we have evidence that (1) the ensemble used throughout the paper is robust and (2) reflects \emph{all} of Wisconsin's stated redistricting criteria according to the state constitution.

Lastly, we considered alternative definitions of the summary
statistics $H$, $L_\rep$, and $L_\dem$. Instead of shifting the
election data so the resulting global elections margins varied between
45\% and 55\% on might want to take a symmetric interval around the
actual  global elections margins. Taking a range of $\pm7.5\%$ for the
shift, we
produced a second set  statistics: $\widetilde H$, $\widetilde L_\rep$, and $\widetilde L_\dem$.
\begin{table}[h]
  \centering
  \begin{tabular}{l | lll}
    & $\widetilde H$ & $\widetilde L_\rep$ & $\widetilde L_\dem$\\
    \hline 
    WSA12 &100\% & 99.593\% & 73.718\%\\
    WSA14 & 99.99\%& 99.505\%& 91.232\%\\
    WSA16 &97.795\%& 99.004\% & 87.484\% \\
  \end{tabular}  
\end{table}
Again we see that the Wisconsin's plan is still an extreme outlier. The
only change is that the $\widetilde L_\rep$ statistic is much
higher. As we discuss bellow, this is because the range now includes a
range of percentages where the Wisconsin plan causes the Democrats to
perform better than expected in the typical plan. However the results
in this range have little effect on the balance of power as the
Republicans are already solidly in the majority in those elections.

We prefer $H$, $L_\rep$, and 
$L_\dem$ to  $\widetilde H$, $\widetilde L_\rep$, and 
$\widetilde L_\dem$ because the range is limited to  45\% to 55\%. 
While the others are more symmetric, they often pull information from 
the low 60\% or high 30\% in global vote. These ranges seem less 
relevant.  The effect of this difference is seen in the values of 
$\widetilde L_\dem$ which is much higher than $L_\dem$ because in 
includes elections with a large global percentage of Republican 
votes. From Figure~\ref{fig:boxPlots}, we see that the Democratic 
votes depleted from districts with partisen make up around 50\% often 
is packed into districts with more that 60\%. This causes a tilt 
in favore of   the Democrats from what is expected should the global 
vote get that high. Of course if the vote is above 60\% Republican, a 
few seats shifted to the Democrats will have little effect 
operationally.
\section{Adjustments to  Wisconsin General Assembly Redistricting}\label{sec:adjust}
Data provided in \cite{WIShapeFileData} is incomplete in terms of the
current redistricting plan for Wisconsin.  We provide the script that
we used to assign districts to unreported wards in our
repository. The number of wards affected is relatively small.

\section{Supplementary Materials}

Database with redistricting plans and other data: \\
\url{git@git.math.duke.edu:gjh/WIRedistrictingData.git}

\section{Acknowledgements} This work uses a code base initiated by Han
Sung Kang and Justin Luo as part of a Data+ project under the
supervision of the authors at Duke University. We thank the Information Initiative at Duke and
the Mathematics Department for their support.
We would also like to thank Moon Duchin, Assaf Bar-Natan, and Mira
Bernstein for their guidance on districting criteria in Wisconsin and
assistance with gathering and extracting data. We are also indebted to
Eric Lander for useful discussions and debates around the 
meaning and presentation of these results as well as Jordan
Ellenberg's insightful comments on a previous draft.  We are also indebted to
Venessa Barnett-Loro for helping to polish this report. 

\bibliography{gerrymandering}

\end{document}